\documentclass[aps,twocolumn,floats,longbibliography,superscriptaddress]{revtex4-2}
\usepackage{graphics,graphicx,epsfig}
\usepackage{amssymb,color}
\usepackage{epsf,epstopdf}
\usepackage {amsmath}
\usepackage[final,hidelinks]{hyperref}
\usepackage{ragged2e}

\newcommand{\abs}[1]{|#1|}

\newcommand{\beq}{\begin{equation}}
\newcommand{\eeq}{\end{equation}}
\newcommand{\beqn}{\begin{eqnarray}}
\newcommand{\eeqn}{\end{eqnarray}}

\usepackage{comment}
\newcommand{\expect}[1]{ \langle #1 \rangle} 

\usepackage{xstring,etoolbox}

\newcommand{\fixsplit}[2]{\StrLen{#2}[\mynum]\ifnumcomp{\mynum}{<}{\numexpr(#1)+1\relax}%
  {#2}%
  {\StrSplit{#2}{#1}{\myfirststr}{\mysecondstr}\myfirststr\linebreak
  \fixsplit{#1}{\mysecondstr}}}

\begin{document}
\title{Transcription factor clusters as information transfer agents}
\author{Rahul Munshi}
\affiliation{Joseph Henry Laboratories of Physics, Princeton University, Princeton, NJ 08544, USA}
\affiliation{Lewis-Sigler Institute for Integrative Genomics, Princeton University, Princeton, NJ 08544, USA}
\author{Jia Ling}
\affiliation{Lewis-Sigler Institute for Integrative Genomics, Princeton University, Princeton, NJ 08544, USA}
\author{Sergey Ryabichko}
\affiliation{Lewis-Sigler Institute for Integrative Genomics, Princeton University, Princeton, NJ 08544, USA}
\author{Eric Wieschaus}
\affiliation{Lewis-Sigler Institute for Integrative Genomics, Princeton University, Princeton, NJ 08544, USA}
\affiliation{Department of Molecular Biology and Howard Hughes Medical Institute, Princeton University, Princeton, NJ 08544, USA}
\author{Thomas Gregor}
\email[Correspondence: ]{tg2@princeton.edu}
\affiliation{Joseph Henry Laboratories of Physics, Princeton University, Princeton, NJ 08544, USA}
\affiliation{Lewis-Sigler Institute for Integrative Genomics, Princeton University, Princeton, NJ 08544, USA}
\affiliation{Department of Stem Cell and Developmental Biology, CNRS UMR3738 Paris Cité, Institut Pasteur, 25 rue du Docteur Roux, 75015 Paris, France}

\date{\today}

\begin{abstract}
Deciphering how genes interpret information from transcription factor (TFs) concentrations within the cell nucleus remains a fundamental question in gene regulation. Recent advancements have revealed the heterogeneous distribution of TF molecules, posing challenges to precisely decoding concentration signals. Using high-resolution single-cell imaging of the fluorescently tagged TF Bicoid in living \textit{Drosophila} embryos, we show that Bicoid accumulation in submicron clusters preserves the spatial information of the maternal Bicoid gradient. These clusters provide precise spatial cues through intensity, size, and frequency.  We further discover that gene targets of Bicoid, such as Hunchback and Eve, colocalize with these clusters in an enhancer binding affinity-dependent manner. Our modeling suggests that clustering offers a faster sensing mechanism for global nuclear concentrations than freely diffusing TF molecules detected by simple enhancers.
\end{abstract}
\maketitle

\begin{figure*}
\centering
\includegraphics[width=0.9\linewidth]{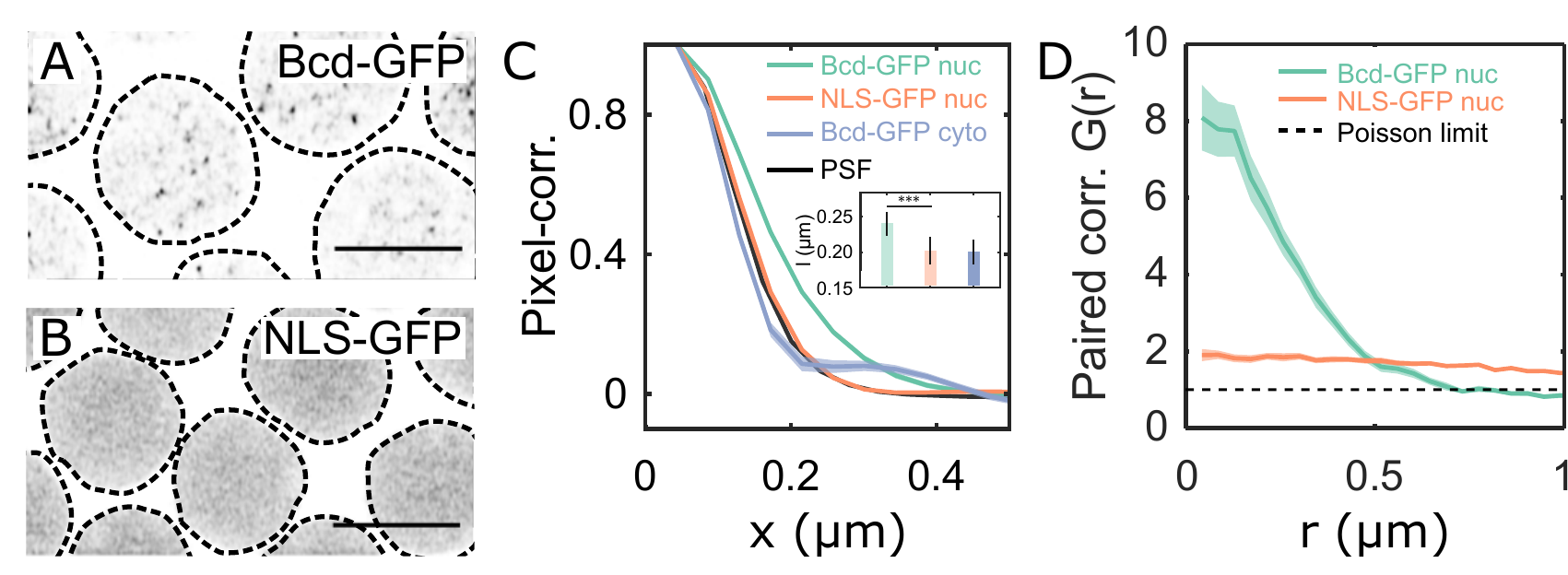}
\caption{{\bf Quantitative characterization of nuclear Bcd heterogeneity.}
(A-B) Confocal (\textit{Zeiss-Airyscan}) images of cross-sections of Bcd-GFP (A), and NLS-GFP (B) expressing blastoderm nuclei in living \textit{Drosophila} embryos (NC14). Scale bars are $5\:\mu$m.
The broken lines represent a guide to the eye for nuclear boundaries.
(C) Pixel correlations computed on the nuclear pixels in 2D nuclear cross-section images (see Materials and Methods) expressing Bcd-GFP (green, $44$ nuclei from $5$ embryos) and NLS-GFP (orange, $27$ nuclei from $3$ embryos); and from pixels within the cytoplasm of Bcd-GFP expressing embryos (grey, $5$ embryos).  For comparison, the objective's point-spread-function (PSF) is in black.
Inset shows mean and standard deviations of the computed correlation lengths ($l$) for nucleoplasmic Bcd-GFP ($l=0.24 \pm 0.02\:\mu m$), nucleoplasmic NLS-GFP ($l=0.20 \pm0.02\:\mu m$), and cytoplasmic Bcd-GFP ($l=0.20 \pm0.02\:\mu$m).
(D) Radial distribution function (or pair-correlation function, $G(r)$) for the local maxima distribution expressed as a function of distance $r$ from the center. $G(r)$ was calculated on time-projected ($60$ frames each) local intensity maxima centroid maps (see Materials and Methods, and Fig.~\ref{figS1D}), averaged over multiple nuclei (same nuclei and embryo count as in C). A distinct peak in G(r) indicates temporally persistent confinement of the local maxima, as seen for Bcd-GFP-expressing nuclei. For NLS-GFP, the continuous reduction in the radial function suggests a gradual decline in intensity near the nuclear edges without sub-micron accumulations. The dashed line ($G(r)=1$) corresponds to a perfectly uniform distribution, the Poisson limit.
}
\label{fig1}
\end{figure*}
%
%
%
%

\section*{\label{Intro}Introduction}
Transcription factors (TFs) play a pivotal role in regulating gene expression by interacting with DNA regulatory elements known as enhancers \cite{spitz2012transcription, 10.1186/s13059-021-02322-1, 10.1016/j.coisb.2022.100438}. These enhancers often exhibit concentration-dependent behavior, activating or repressing gene expression only within specific TF concentration thresholds \cite{10.1126/science.1106914, 10.7554/elife.41266}. The remarkable sensitivity of enhancers to subtle variations in the nuclear concentration of TF molecules implies that genes and enhancers carry out precise measurements of TF concentration \cite{10.1007/s00018-010-0536-y, 10.1016/j.devcel.2023.10.001}. 
\par
However, the challenge arises because TF levels are often quite low and TF molecules are not uniformly distributed in the nucleus \cite{10.1242/jcs.110.15.1781}. 
Instead, they assemble into dynamic transcriptional microenvironments called transcriptional hubs \cite{10.1038/s41467-018-07613-z, 10.7554/elife.45325, 10.1038/s41467-023-40485-6}. These TF molecule accumulations are believed to form through transient clustering mechanisms \cite{10.7554/elife.27451, 10.15252/embj.2018100809, 10.1093/nar/gkad227, 10.1016/j.molcel.2023.04.018} or through liquid-liquid phase separations (LLPS) , \cite{10.1016/j.cell.2018.10.042, 10.1016/j.cell.2017.02.007, 10.1038/s41556-020-00578-6}. Separation of LLPS clusters reflects saturation kinetics, such that increasing concentration of the minor component results in increased size of droplets rather than an increase in the concentration within droplets \cite{10.1016/j.molcel.2019.07.009, 10.1038/s41573-022-00505-4}. Whether droplet size provides a useful proxy for global nuclear concentration is unclear.
\par
In this study, we aim to investigate whether the physical features of these TF assemblies such as size, concentration, or total molecular content accurately reflect the nuclear concentration. We leverage the unique characteristics of the \textit{Drosophila} TF Bicoid (Bcd), known for its varying concentration along the anterior-posterior (AP) axis of the early embryo \cite{10.7554/elife.28275}. Despite low nuclear concentrations, Bcd exhibits an extraordinarily reproducible profile, revealing precision in positional information comparable to the size of a single cell \cite{10.1016/j.cub.2014.04.028, 10.1534/genetics.114.171850, 10.1016/j.cell.2018.09.056}.
\par
Various imaging approaches have unveiled that, similar to many other TFs, Bcd is not homogeneously distributed in the nucleus \cite{abu2010high, 10.1242/dev.202128, 10.1101/2024.01.30.578077, 10.1101/gad.305078.117}. Instead, it forms numerous cluster-like droplets enriched with chromatin accessibility factors like Zelda  \cite{10.1038/s41467-018-07613-z} and actively transcribed canonical Bcd target genes, such as Hunchback, \cite{mir2018dynamic}. The higher concentrations within Bcd accumulations are believed to enhance transcription by increasing the local concentration near target enhancers \cite{10.1038/ncomms8445, 10.1016/j.molcel.2023.04.018}. However, for these clusters to be functionally relevant to Bcd's well-characterized role in patterning, some features of the observed clusters must convey positional information with a precision similar to the nuclear concentration profile.  
\par
Here we developed a quantitative imaging strategy to decipher which features of Bcd accumulations maintain information about concentration. Contrary to simple LLPS models, we found that cluster size remains independent of concentration, while the cluster concentration varies linearly with nuclear Bcd concentration. These clusters localize at the locus of active target genes, precisely conferring information about cellular position. We use these data to quantitatively explore the impact of clustering on information transfer and discuss the circumstances where clustering might be a preferred mechanism as opposed to the gene interacting with the TF molecules freely diffusing in the nucleus.
%
%
%
%


\section*{\label{Res}Results}

\noindent\textbf{Heterogeneity of nuclear TF distribution.}
We revisit the heterogeneous distribution of Bicoid (Bcd) within nuclei to establish quantitative insights. All data presented in this study is derived from live samples unless stated otherwise. The distribution of Bcd within the nucleus comprises both freely diffusing molecules in intranuclear spaces and those engaging with chromatin \cite{10.7554/elife.28275, 10.1101/gad.234534.113}. Cross-sectional images of Bcd-GFP nuclei ($1\:\mu m$ thick z-section) revealed multiple focal accumulations per cross-section (Fig.~\ref{fig1}A, Fig.~\ref{figS1A}A-D and Fig.~\ref{figS1E}). Conversely, embryos expressing an NLS-GFP fusion construct, where molecules diffuse freely without chromatin interaction, showed no such heterogeneity (Fig.~\ref{fig1}B). 

Quantitative analysis of the focal accumulations' average sizes in these cross-sectional images, determined by pixel correlation functions, revealed an average correlation length of $240 \pm 20\:$nm for nuclear Bcd-GFP. In contrast, NLS-GFP expressing nuclei exhibited a smaller correlation length of $200 \pm 20\:$nm, comparable to cytoplasmic Bcd-GFP ($200 \pm20\:$nm) (Fig.~\ref{fig1}C). Both, nuclear NLS-GFP and cytoplasmic Bcd-GFP molecules are freely diffusing, and hence their correlation functions coincide with the microscope objective's point spread function (PSF, Fig.~\ref{figS1A}F). Nuclear Bcd-GFP however, forms focal accumulations larger than the diffraction limit.

\begin{figure*}
\centering
\includegraphics[width=0.75\paperwidth]{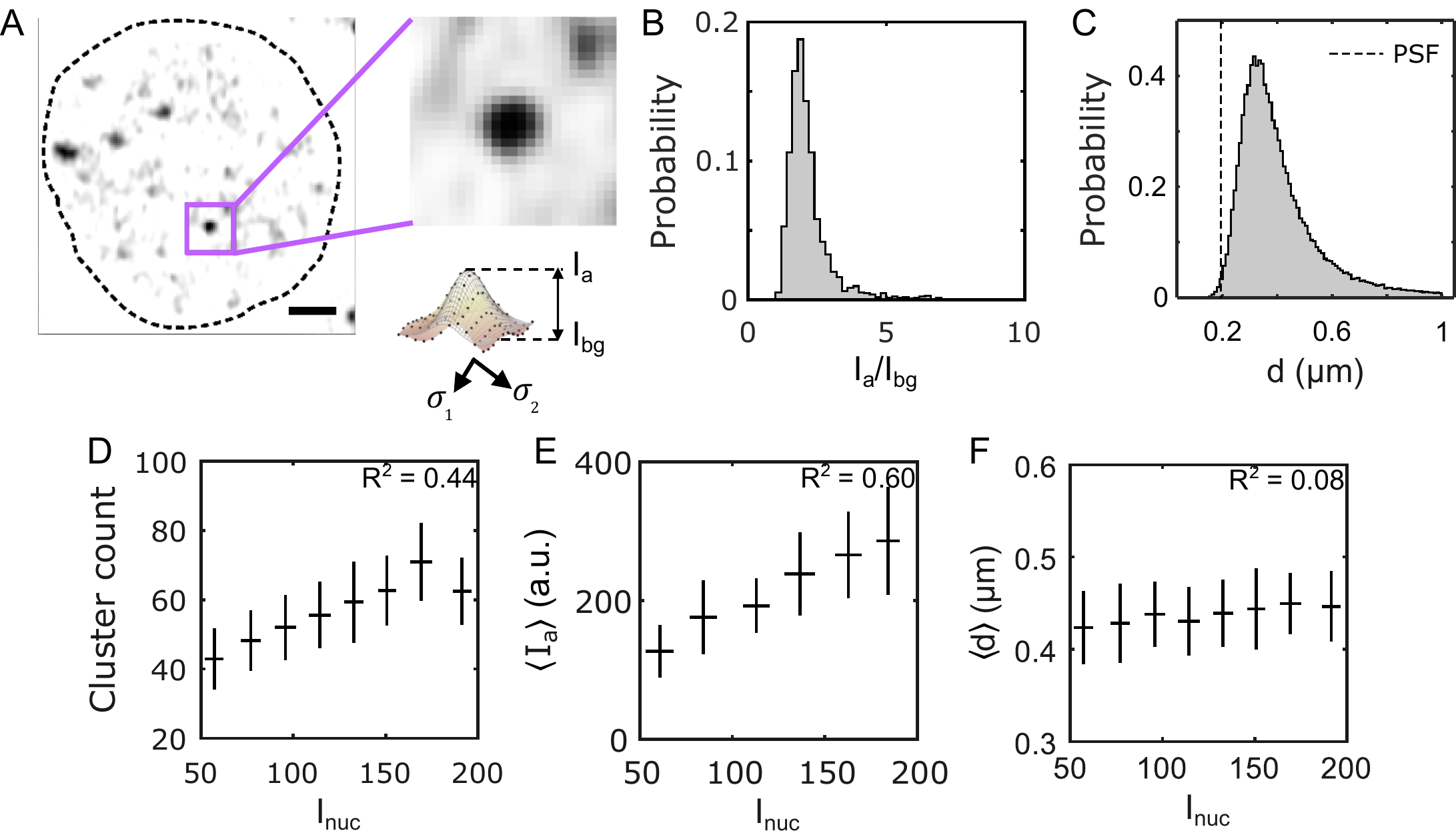}
\caption{{\bf Biophysical properties of Bcd clusters} 
(A) A single nucleus showing Bcd-GFP heterogeneities. The close-up image (right) shows a single Bcd-GFP cluster. Cluster intensity fit with a 2D Gaussian (see profile below). The cluster amplitude ($I_a$), the cluster background intensity ($I_\mathrm{bg}$), and the cluster size ($d$) are extracted from fit parameters (Materials and Methods). Scale bar is $1\,\mu$m.
(B) A histogram of the signal-to-background ratio ($I_a/I_\mathrm{bg}$) for $99671$ clusters from $2027$ nuclei in $14$ embryos expressing Bcd-GFP is plotted.
(C) A histogram of the cluster size ($d$), computed from the same clusters as in B is shown. The vertical dashed line representing the size of the PSF is included to compare with the size of the detected clusters.
(D, E, F) The number of clusters per nucleus (D), the nuclear average of cluster amplitude $\expect{I_a}$ (E), and the nuclear average of cluster size $\expect{d}$ (F) are plotted against nuclear Bcd-GFP intensity, $I_{nuc}$. Error bars represent the mean $\pm$ s.d. for data in each $I_{nuc}$ bin, calculated via bootstrap sampling of data within each bin. The coefficient of determination for each plot in D, E, and F is indicated in the respective panels.
}
\label{fig2}
\end{figure*}

We took short videos ($30$ s long) of nuclear cross-sections to investigate if the focal Bcd accumulations are spatiotemporally persistent. Local GFP fluorescence intensity maxima were identified in each video frame (Materials and Methods), and all frames were combined to form projection maps of local maxima (Fig.~\ref{figS1D}D). The projection maps revealed that the Bcd-GFP maxima tend to crowd inside confinement areas within the nucleus, contrasting with the dispersed maxima in NLS-GFP nuclei (Fig.~\ref{figS1D}A, B, E). Pair-correlation analysis \cite{10.1371/journal.pone.0031457} indicated an effective radius ($\xi_{pair}$) of the confinement area for Bcd-GFP nuclei as $370\pm50\:$nm (Fig.~\ref{fig1}D). No such correlation was detected in NLS-GFP nuclei.
The density of local maxima is approximately eight times higher in the confinement areas compared to the rest of the nucleus (Fig.~\ref{figS1D}F, G). This increase is due to the persistent localization of maxima into sub-micron spaces during the imaging period, which is suggestive of clustering. 

The cluster lifetime and the frequency of cluster formation can adequately describe clusters' temporal persistence. We calculated the cluster lifetime ($T_{on}$) and the inverse of the cluster frequency ($T_{off}$) from the maxima in the confinement area and found the effective $T_{on}$ and $T_{off}$ to be $2.4\pm0.3$ s and $1.6\pm0.3$ s, respectively (Fig.~\ref{figS1D}H). The fraction $T_{on}/(T_{on}+T_{off})$ gives the probability of cluster detection, which was found to be $58\:\%$ (Fig.~\ref{figS1D}I). These results indicate that Bcd accumulations form persistent sub-micron clusters within the nucleus. To assess the potential of these clusters in transferring information to target genes, we proceed to characterize their biophysical properties.   

\begin{figure}[t]
\centering
\includegraphics[width=\linewidth]{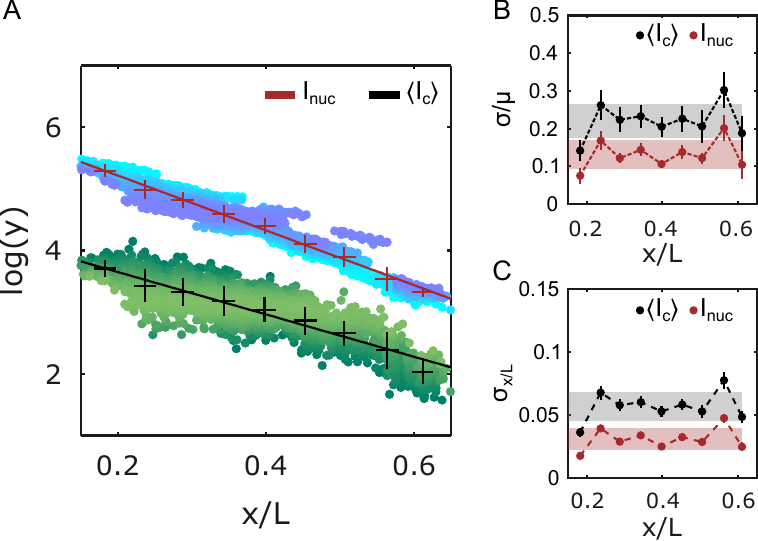}
\caption{{\bf Precision of cluster positional information potential.} 
(A) Overall Bcd-GFP nuclear intensity ($I_\mathrm{nuc}$) and the nuclear average of Bcd-GFP cluster intensity ($\expect{I_c}$) as a function of nuclear position  $x/L$ (with embryo length $L$). $\expect{I_c}$ measures the molecular count within the clusters (Fig.~\ref{figS3C} and  Materials and Methods). Y-axis is in natural logarithm units. Blue ($I_\mathrm{nuc}$) and green ($I_\mathrm{c}$) shaded data points represent individual nuclei ($2027$ nuclei in $14$ embryos). Data is partitioned in $x/L$-bins (mean and s.d. shown, error bars calculated from bootstrapping; exponential decay constants extracted from linear fits (solid lines) with $\lambda_{I_\mathrm{nuc}}=0.23\pm0.03\,L$, and $\lambda_{I_c}=0.26\pm0.02\,L$). 
(B) Coefficients of variation (c.v.) ($\sigma/\mu$) for $I_\mathrm{nuc}$ and $\expect{I_c}$ as a function of $x/L$-bins. 
(C) Errors in determination of nuclear positions using $I_\mathrm{nuc}$ (red) and $\expect{I_c}$ (grey) as a function of $x/L$-bins (obtained via error propagation, Materials and Methods). For B and C, grey and red shades indicate the overall mean $\pm$ s.d. across all positions for $\expect{I_c}$ and $I_{nuc}$, respectively.
}
\label{fig3}
\end{figure}

\begin{figure*}[t]
\centering
\includegraphics[width=0.65\paperwidth]{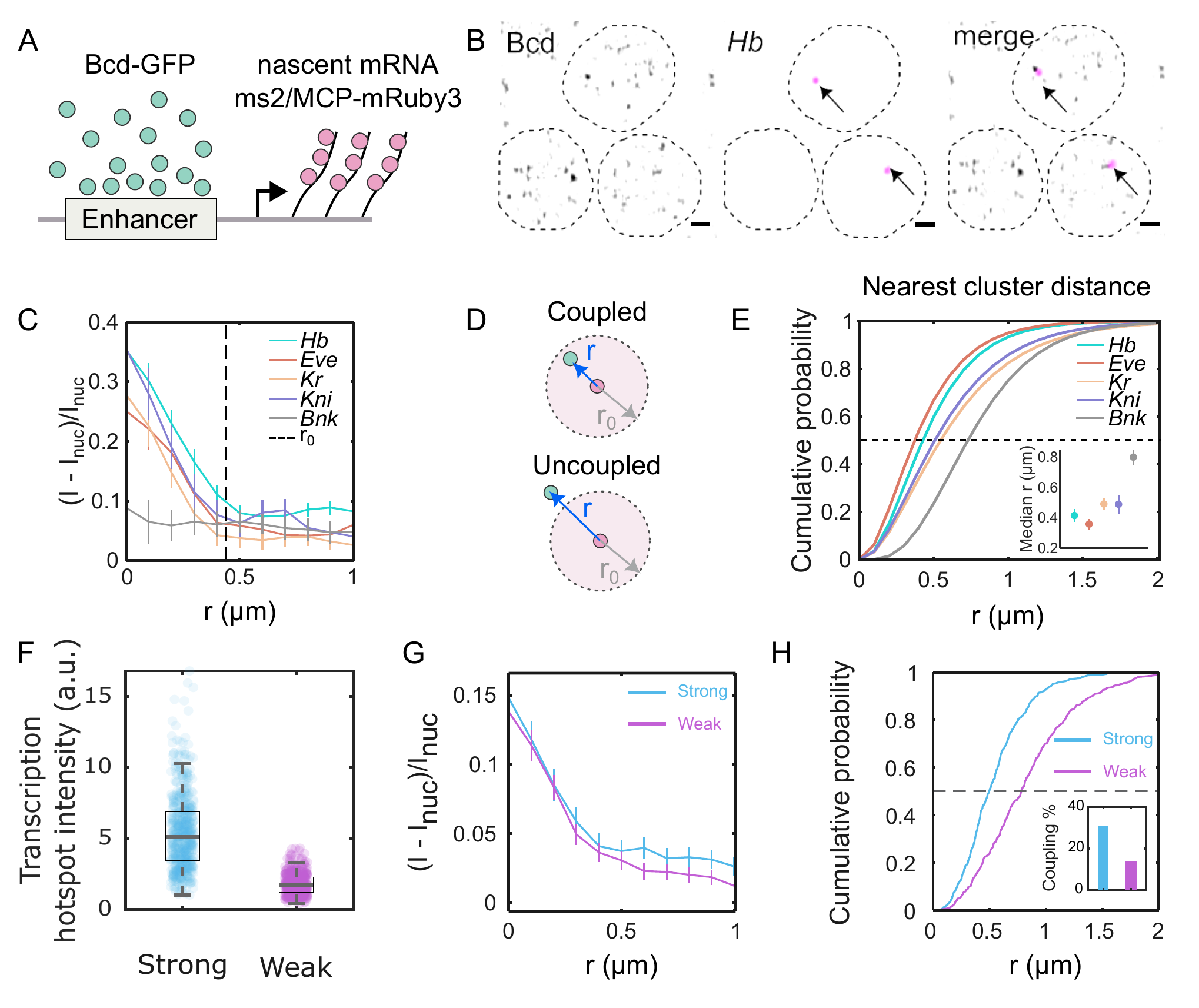}
\caption{{\bf Bcd cluster co-localization with target genes is enhancer dependent.} 
(A) Cartoon showing scheme for dual color imaging with Bcd-GFP (green) and nascent transcription site labeled via the MS2/MCP system (magenta). 
(B) Images from embryos in NC14 showing nuclei expressing Bcd-GFP and  \textit{hb}-MS2/MCP-mRuby on sites of active transcription (arrows); scale bar is $1\,\mu$m. Dashed lines are a guide to the eye for nuclear boundaries. 
(C) Radial distribution of Bcd-GFP intensity around the centroid of the fluorescently labeled gene locus (i.e. hotspot). Data shown for canonical Bcd target genes, \emph{hb} (102 nuclei, 13 embryos), \emph{eve} (66 nuclei, 8 embryos), \emph{Kr} (107 nuclei, 11 embryos), \emph{kni} (90 nuclei, 6 embryos), and the non-target gene \emph{bnk} (56 nuclei, 10 embryos). Dashed line ($r_0=0.44\pm 0.05\:\mu m$) is twice the full width at half maximum (FWHM) averaged over all genes. Data is obtained from simultaneous imaging of Bcd-GFP and MCP-mRuby3, marking the nascent transcription hotspots of the respective genes (Materials and Methods).
(D) Schematic showing the mRNA hotspot (red) and its nearest Bcd cluster (green). When the distance $r$ between the nearest cluster and the hotspot is less than the Bcd accumulation radius $r_0$, the cluster is defined as being \textit{coupled} to the gene; when it is greater than $r_0$ the cluster is assumed to be \textit{uncoupled} (see also Fig.~\ref{figS4B}). 
(E) Cumulative probability distributions of distances ($r$) between the mRNA hotspot and its nearest cluster, computed for the same data as in C. Dashed line is the median at EC50. Inset: Median distances for all genes. Errors are calculated from bootstrapping.
(F) Distributions of transcription hotspot intensities from a synthetic strong (blue, 541 nuclei, 17 embryos) and weak (magenta, 406 nuclei, 20 embryos) enhancer constructs driving an MS2-fusion reporter (see Materials and Methods and Fig.~\ref{figS4C}). The strong construct generates a 3.2-fold higher intensity than the weak construct, on average. Boxes represent the 1\textsuperscript{st} and the 3\textsuperscript{rd} quartiles, while the whiskers represent the 5\textsuperscript{th} and the 95\textsuperscript{th} percentiles. The medians (black lines inside the boxes) are 5.1 and 1.7 for the strong and the weak enhancers, respectively.
(G) Radial distributions of relative Bcd-GFP intensities around the centroid of the transcription hotspot. 
The accumulation radii are statistically identical ($0.36\pm0.05\:\mu $m and $0.39\pm0.06\:\mu$m for strong and weak enhancer constructs, respectively).
(H) Cumulative probability distributions of distances $r$ between the transcription hotspot and its nearest Bcd cluster. The black dashed line is at EC50. The median distances are $0.49\pm0.03\:\mu$m and $0.78\pm0.05\:\mu$m for the strong and weak constructs, respectively.
Inset shows the fraction of coupled clusters for each construct ($31\:\%$ and $13\:\%$, respectively).
}
\label{fig4}
\end{figure*}

\vspace{.3 cm}
\noindent\textbf{Bcd cluster properties.}
To define the extent of Bcd clusters in 3D and characterize their biophysical properties, we took an approach different from the maxima detection approach used in the previous section. Since any cluster should be at least the size of the PSF along the x-y plane ($>4$ pixels), an x-y cutoff of 3 pixels should eliminate any spurious spots. Also, from the persistence data (Fig.~\ref{figS1D}H), we argued that for an imaging frame time of $\sim500\:$ms, a cluster should span across at least two consecutive z frames, given the frame thickness is less than the z PSF (Materials and Methods). Using this approach, between 40 and 70 clusters could be identified per nucleus.

Individual cluster parameters are determined from 2D Gaussian fits of the GFP intensity profile at the z-plane of the cluster centroid (Fig.~\ref{fig2}A). These fits estimate the effective diameter $d$, providing a measure for the cluster size (see Materials and Methods). The average cluster size per nucleus is $\expect{d}= 400\pm140\:$nm for all clusters (Fig.~\ref{fig2}C). Notably, the left tail of the cluster size distribution vanishes around the PSF limit, despite applying a considerably smaller size cutoff ($150\:$nm) than this limit (Fig.~\ref{fig2}C). This suggests that the detectable clusters are not diffraction-limited under our imaging conditions. Sub-diffraction clusters may exist with very low intensities that evade detection or are highly transient, making them undetectable within the scope of this study.

To gauge the cluster concentration, we introduce the parameter $I_\mathrm{a}$, representing the peak cluster intensity, or cluster amplitude, and $I_\mathrm{bg}$, denoting the concentration of Bcd molecules in the nuclear space surrounding the cluster (Fig.~\ref{fig2}A and Materials and Methods). The signal-to-background ratio $I_\mathrm{a}/I_\mathrm{bg}$ offers insights into local Bcd concentration amplification within a cluster, with an average value of $2.2\pm0.8$ for close to $10^5$ clusters (Fig.~\ref{fig2}B).

Since the nuclear Bcd concentration (given by $I_\mathrm{nuc}$) varies exponentially along the embryo axis, we sought to understand how the cluster properties change with concentration. The cluster count showed a strong dependence on $I_\mathrm{nuc}$, exhibiting an almost two-fold drop across the anterior $\sim60\:\%$ of the embryo length (Fig.~\ref{fig2}D, also see Fig.~\ref{figS2B}). This indicates that clustering occurs less frequently in nuclei with lower Bcd concentration. 

We found that $\expect{I_\mathrm{a}}$ also shows a strong dependence on $I_\mathrm{nuc}$, with an almost two-fold change within $\sim60\%$ of the embryo (Fig.~\ref{fig2}E), while $\expect{d}$ varies only insignificantly (Fig.~\ref{fig2}F, and Fig.~\ref{figS3B}G). The distribution of the mean and variance of $d$ remained similar at various ranges of $I_\mathrm{nuc}$ (Fig.~\ref{figS2C}B, C). Furthermore, there was no correlation between $d$ and $I_\mathrm{a}$ (Fig.~\ref{figS2C}A). This led us to conclude that the cluster size $d$ is independent of both the nuclear concentration ($I_\mathrm{nuc}$) and cluster concentration ($I_\mathrm{a}$) of Bcd.

Thus, one might speculate that droplet growth by coalescence at higher concentrations, a characteristic of LLPS condensates, might be absent in Bcd clusters \cite{10.1016/j.tcb.2019.10.006}. This speculation is complemented by the observation that the dependence of the molecular concentration of an average cluster (given by $\expect{I_\mathrm{a}}$) on the nuclear concentration ($I_\mathrm{nuc}$) is approximately linear ($R^2=0.6$) (Fig.~\ref{fig2}E, and Fig.~\ref{figS3B}D). This linearity contrasts with the switch-like dependence observed in LLPS condensates \cite{10.1038/s41573-022-00505-4}. Notably, we observe clustering even in nuclei with very low Bcd concentrations, indicating the absence of a discernible threshold concentration triggering cluster formation \cite{10.1038/s41586-021-03905-5}. However, further investigation is warranted to ascertain whether the clusters analyzed represent a matured molecular state where conventional LLPS rules no longer apply, or if detailed imaging, capturing cluster formation dynamics is needed to distinguish between these possibilities.

\vspace{.3 cm}
\noindent\textbf{Do clusters contain enough positional information?} 
Previously, we have established that the position of anterior nuclei in the early \textit{Drosophila} embryo can be determined with a spatial precision of better than $1\:\%$ from nuclear Bcd concentration alone \cite{10.1016/j.cell.2007.05.025}. This precision stems from the collective contribution of all nuclear Bcd molecules reproducible to within $10\:\%$. Given that clusters comprise only a small fraction of molecules within nuclei (Fig.~\ref{figS3C}B), we sought to investigate whether they could offer an accurate estimation of nuclear concentration and, consequently, the position of the nucleus along the AP axis of the embryo.

To this end, we consider the cluster intensity, $I_c$, a representative of the molecular count of Bcd within a cluster, such that $I_c = 2I_a\sigma_1\sigma_2$, where $\sigma_1$ and $\sigma_2$ are characteristic cluster fit parameters (see Fig.~\ref{fig2}A and Materials and Methods). From this quantity, an absolute count for the total Bcd molecules within a cluster can be computed using previous estimates for absolute molecular count conversions \cite{10.7554/elife.28275} (Fig.~\ref{figS3C}C).

The nuclear average of $\expect{I_c}$, decays exponentially with the nuclear position like the $I_{nuc}$ gradient: the exponential decay constants of $\expect{I_c}$ and $I_{nuc}$ are statistically very similar (Fig.~\ref{fig3}A).  Thus, the molecular count of an average cluster mirrors the Bcd nuclear concentration gradient. The corresponding average coefficients of variation (Fig.~\ref{fig3}B) for $I_\mathrm{nuc}$ and $\expect{I_c}$ are $14\pm4\:\%$ and $22\pm4\:\%$, respectively. Therefore, despite the clusters representing only a small fraction of nuclear Bcd molecules ($5-10\:\%$, Fig.~\ref{figS3C}), the $\expect{I_c}$-derived Bcd gradient displays remarkably low variability, hinting at the existence of tightly controlled mechanisms that regulate cluster formation.

As a morphogen, Bcd's nuclear concentration confers positional identity to a nucleus with sufficiently high accuracy, such that neighboring nuclei in the anterior $60\:\%$ can be distinguished by reading the nuclear Bcd concentration alone \cite{10.1016/j.cell.2007.05.025, liu2013dynamic}. To estimate the level of positional information contained in the cluster-derived Bcd gradient, we estimate the error in position determination $\sigma (x)$ from the Bcd gradient's concentration fluctuations $\delta c(x)$ \cite{10.1016/j.cell.2007.05.025}. Using simple error propagation, the error in position determination is given by $\sigma(x) = \delta c(x) |\frac{c(x)}{dx}|^{-1}$, where $c(x)$ is the Bcd concentration at position $x$.

Using $\expect{I_c}$ as the estimator for nuclear Bcd concentration, the error in position determination (i.e., the positional error) is $\sigma(x)=5.5\pm0.7\:\%\:L$ (Fig.~\ref{fig3}C). This corresponds to a positional precision of roughly three cell diameters, significantly less precise than the single-cell precision previously shown with the full nuclear Bcd concentration $I_\mathrm{nuc}$. Similarly, $\expect{I_c}$ can also be used to estimate $I_{nuc}$. In that case, near the anterior of the embryo, the error is $\sim15\:\%$ (Fig.~\ref{figS3B}C), which is comparable to the variability of $I_{nuc}$ itself (Fig.~\ref{fig3}B), making the average nuclear cluster concentration a very good proxy for the overall nuclear Bcd concentration. The errors in nuclear concentration determination as well as nuclear position determination were computed for other cluster properties, such as $\expect{I_a}$ and $\expect{d}$, but in this case, the errors were found to be higher than those using $\expect{I_c}$ (Fig.~\ref{figS3A}, \ref{figS3B}). 

The estimation obtained from $\expect{I_c}$ reflects the property of an average cluster. Individual clusters might confer positional information with varying accuracy, with the highest potentially being equivalent to $I_\mathrm{nuc}$. However, for genes to utilize this information, the clusters must be physically close to specific gene loci, which we examine next. 

\vspace{.3 cm}
\noindent\textbf{Cluster association with target genes.}
To elucidate the behavior of individual clusters around target gene transcription sites, we conducted three-dimensional imaging of labeled nascent mRNAs of putative target genes (\emph{hunchback}, \emph{even-skipped}, \emph{Krüppel}, \emph{knirps}) \cite{10.1038/340363a0, 10.1038/287795a0} while imaging Bcd-GFP within the nuclei (Fig.~\ref{fig4}A, B, and Fig.~\ref{figS4A}). For each of the target genes, Bcd accumulation was observed, with Bcd-GFP intensity peaks at the center of the nascent mRNA hotspot (Fig.~\ref{fig4}C). The radii of Bcd-GFP accumulation around the four target genes were determined to be $490\pm40\:nm$, $550\pm20 \:nm$, $390\pm40\:nm$, and $330\pm30\:nm$ for \emph{hunchback}, \emph{even-skipped}, \emph{Krüppel}, \emph{knirps}, respectively (Fig.~\ref{figS4C}A). These radii were comparable to the radius of the average enrichment area shown in Fig.~\ref{fig1}D (see Fig.~\ref{figS4B}A for a simulation-based representation).

In contrast, Bcd accumulation was not detected around a non-target gene, \emph{bottleneck} \cite{10.1016/0092-8674(93)80078-s} (Fig.~\ref{fig4}C). Nor does Bcd accumulate around the geometric nuclear centers, considered random sites unassociated with a particular gene locus (Fig.~\ref{figS4B}B). Specificity was confirmed by imaging NLS-GFP in place of Bcd-GFP, revealing no accumulation around the \emph{hunchback} locus (Fig.~\ref{figS4B}B).

The presence of Bcd accumulation near target gene loci indicates that Bcd clusters tend to have a high probability of colocalizing with the gene loci. However, a TF cluster may not be directly associated with the gene locus throughout the entire duration of active transcription of the gene locus. In such cases, the nearest TF cluster would be uncoupled from the gene (Fig.~\ref{fig4}D), leading to a greater physical distance from the gene transcription site than a coupled cluster. 
The TF accumulation radius (Fig.~\ref{figS4C}A) gives a confinement radius within which a coupled cluster can be located. Utilizing this accumulation radius, a distance limit for cluster-gene coupling can be established, where any TF cluster located within that distance limit can be considered coupled to the respective gene.

The median 3D distances of the nearest Bcd clusters from the center of the genes (mRNA hotspots) were determined to be $420\:$nm, $360\:$nm, $500\:$nm, and $490\:$nm for \emph{hunchback}, \emph{even-skipped}, \emph{Krüppel}, and \emph{knirps}, respectively; and it was $800\:$nm for the non-target gene \emph{bottleneck} (Fig.~\ref{fig4}E). Applying the respective distance limits (Fig.~\ref{fig4}D,  Fig.~\ref{figS4C}A) to the cumulative probability plots of the nearest cluster distance distributions, we calculated the fraction of clusters coupled to the respective genes (an alternate technique yielding similar results is shown in Fig.~\ref{figS4B}C). The fractions of coupled clusters were $0.57$, $0.73$, $0.41$, and $0.30$, respectively, for \emph{hunchback}, \emph{even-skipped}, \emph{Krüppel}, and \emph{knirps} (Fig.~\ref{figS4C}B). Since an accumulation radius is not well-defined for \emph{bottleneck}, no localization fraction could be determined for this gene.

\begin{figure}[t]
\centering
\includegraphics[width=\linewidth]{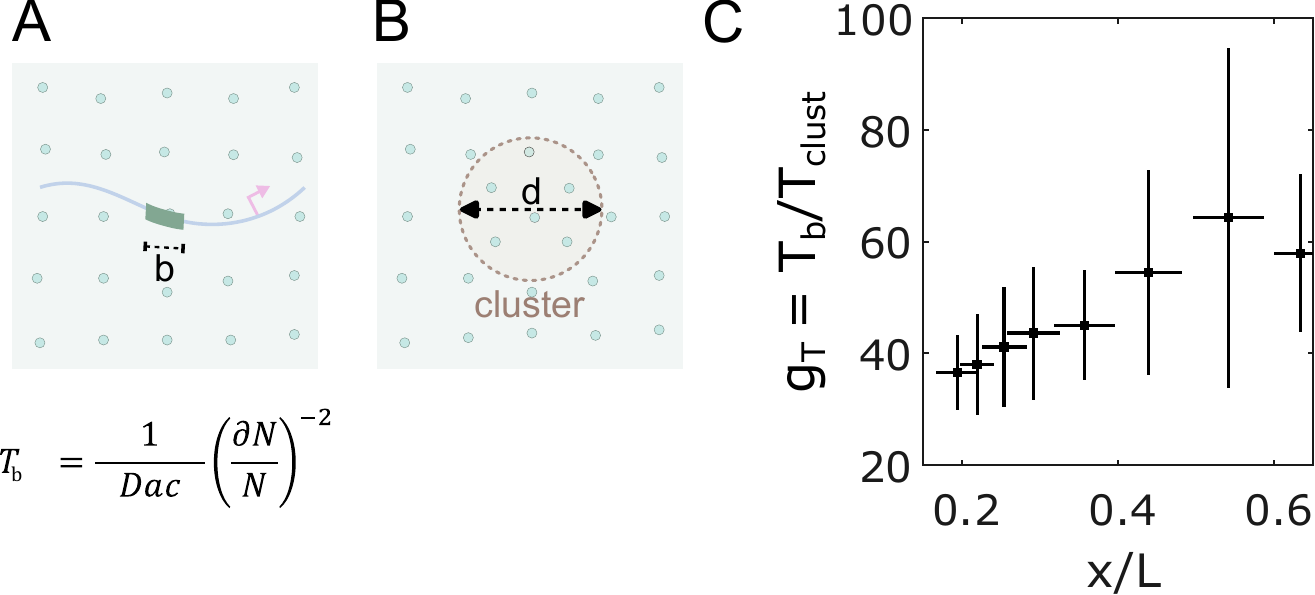}
\caption{{\bf Clustering reduces time to precise concentration interpretation.} 
(A, B) Two cartoons show Bcd molecules in the nucleus (green circles) and an enhancer with a binding site of length $b$ (A) or a cluster of diameter $d$ (B) embedded in the nuclear environment. 
The equation in (A) is for the time taken by a sensor of size $a$ for nuclear concentration $c$ with an accuracy of $\left(\frac{dN}{N}\right)$, where $N$ is the number of molecules counted.
(C) Reduction of time ($g_T$) to make an accurate ($\sim 10\%$) nuclear concentration estimation as a function of the nuclear position with the cluster as nuclear concentration sensor versus an enhancer binding site being the concentration sensor \cite{10.1016/j.cell.2007.05.025}.
}
\label{fig5}
\end{figure}

These findings suggest that Bcd clusters tend to localize with target genes with high probability. Localization is also enhancer-dependent. When comparing strong and weak enhancers for the \emph{hunchback} gene (Fig.~\ref{figS4C}) (see Materials and Methods), the strong enhancer produces a much higher transcriptional output (Fig.~\ref{fig4}F), despite the similar radial average of Bcd concentration at the transcription site for both (Fig.~\ref{fig4}G). However, Bcd clusters are closer to the transcription site for the strong enhancer, which also has a higher fraction of clusters bound to the active target compared to the weak one (Fig.~\ref{fig4}H, inset). Additionally, $I_a$ of the nearest Bcd clusters is more strongly correlated with $I_{nuc}$ for strong enhancers than for weak enhancers (Fig.~\ref{figS4D}).

Given that Bcd clusters carry information about the nuclear position, genes can thus access this information by directly interacting with the clusters. However, genes can also interpret the same information by directly interacting with molecules diffusing in the nucleus. Thus, the question arises: why is clustering favored?

\vspace{.3 cm}
\noindent\textbf{Clusters are fast information sensors.} 
Nuclear Bcd concentration has to be interpreted by Bcd's target gene loci to accurately extract positional information from the morphogen gradient and trigger a transcriptional response accordingly. The ability to sense diffusing TF molecules naturally depends on the effective size of the sensor. Whether the size is of the order of a binding site ($3\,$nm), an enhancer ($50\,$nm), or the entire locus is unclear. We know from previous estimations that if a binding site is a relevant size metric, then the measured readout precision needs to invoke spatial averaging across multiple independent sensors, i.e. neighboring nuclei \cite{10.1016/j.cell.2007.05.025}. 
\par
Here we consider the possibility that the TF cluster functions as a sensor and that the transcriptional output of a gene (its sensing of the gradient) is a reflection of the Bcd content of the cluster rather than interaction between single Bcd molecules at individual target gene enhancers. In previous analyses focused on enhancers, the time required for interpreting nuclear concentration was estimated using a molecular sensing argument \cite{10.1016/S0006-3495(77)85544-6, 10.1016/j.cell.2007.05.025}. To apply this approach to Bcd’s heterogeneous distribution in nuclei, we treated a cluster as a sphere with an effective diameter $d$, the concentration of Bcd molecules inside the cluster as $c_\mathrm{clust}$, and the diffusion constant of Bcd as $D$ (Fig.~\ref{fig5}B). The time $T_\mathrm{clust}$ required for the cluster to precisely mirror the global nuclear concentration (the cluster sensing event that interprets nuclear concentration with an accuracy of  $\left(\frac{\partial N}{N}\right)$) can be compared to the time $T_\mathrm{b}$ required for a simple binding site in an enhancer of linear size $b$, to measure nuclear concentration (Fig.~\ref{fig5}A).

The ratio $T_\mathrm{b}/T_\mathrm{clust}$ yields insight into the comparative sensing times. If clusters function as concentration sensors, the average cluster in an anterior nucleus in the embryo could sense nuclear concentration approximately $37.5\pm5.1$ times faster than a single binding site (Fig.~\ref{fig5}C), owing to the larger sensor size and $\sim2$ fold concentration amplification within a cluster (Fig.~\ref{fig2}B). Hence, an average cluster can interpret nuclear concentration in $\sim3$ minutes, which is the timescale relevant to the activation of the target genes \cite{10.1016/j.cell.2007.05.025, 10.1371/journal.pgen.1007676}. Understanding the mechanistic interactions between Bcd clusters and target genes opens new avenues for future gene regulation and transcriptional control research.
%
%
%
%
%
\section*{\label{sec:dis}Discussion}
In this study, we employed quantitative imaging techniques in live embryos to elucidate the role of subnuclear compartmentalization, particularly clustering, in preserving the information carried by signaling molecules within the cell nucleus. Previous research has shown that TF clusters in various organisms and tissue cultures are spatially associated with transcriptionally active sites of target genes \cite{10.15252/embj.2018100809, 10.1016/j.molcel.2023.04.018}. According to the LLPS model, clustering results in a non-stoichiometric assembly of molecules when the global concentration exceeds a specific threshold \cite{10.1038/s41573-022-00505-4}. This raises questions about how such clusters can maintain information about nuclear concentration. More complex models, such as those involving the seeding of droplets on enhancers, also suggest that clusters do not maintain precise nuclear concentration information \cite{10.1016/j.molcel.2019.07.009}. However, it is well established that the expression of response genes is highly sensitive to the global nuclear concentrations of TFs \cite{10.1016/j.cell.2007.05.025}.

To reconcile this dichotomy, we chose to study Bicoid due to its concentration gradient, which offers two key advantages for analysis. Firstly, its graded concentration can be optically measured with single-cell precision. Secondly, the transcriptional responses of multiple target genes can be dynamically measured in living embryos. Our experiments leverage these features and demonstrate that the intensity and total number of Bcd molecules in clusters effectively preserve nuclear concentration information.

These results prompt two fundamental questions about transcription factor clusters. First, how do these clusters form in a way that retains information about concentration? Second, how are the features of these clusters, which convey concentration information, interpreted? Several observations from our study highlight interesting areas for future research.

Any analysis of cluster formation must take into account that each cluster defines a transcriptional microenvironment integrating multiple interacting components, such as Mediator molecules, chromatin-modifying agents, and PolII. Bcd has been shown to interact with several of these components via its activation domain \cite{burz1998cooperative}. Any one of these components may play a central role during the initial stages of cluster formation. Therefore, the formation and effective size of such clusters might reflect the presence of other, or indeed, all constituent molecules rather than being dependent solely on the concentration of a single molecular species like Bcd. 

Bcd interacts with DNA through its DNA binding motif \cite{10.1016/0092-8674(89)90063-9}, which is present in multiple copies within the enhancers that govern its target gene activity. Bcd’s binding to enhancers might seed cluster formation in ways that do not maintain a direct dependence on its concentration, even though that concentration determines the intensity of Bcd’s accumulation in clusters. Our observation that clusters of finite sizes were present even at very low Bcd concentrations is consistent with previous studies reporting clustering at low concentrations \cite{10.1101/gad.305078.117, 10.7554/elife.28975}. 
This is contrary to what might be expected of classic single-molecule LLPS assemblies \cite{10.1101/gad.305078.117,10.1016/j.molcel.2019.07.009, 10.1038/s41573-022-00505-4}, where frequency and size are expected to depend more directly on absolute concentration. Overall our results suggest that cluster formation is driven by multiple molecular species and highlight the potential role of enhancers in cluster seeding.

A second feature complicating our analysis of cluster formation is their stability and dynamics. Our analysis of frequency and size is limited by the constraints imposed by our imaging conditions, which allow us to detect only clusters that are larger than the diffraction limit and stable for periods exceeding 500 ms. While this does not rule out the existence of smaller, highly transient clusters, it is curious how and why the average cluster size remains invariant with transcription factor concentration.

Techniques such as fluorescence correlation spectroscopy (FCS) \cite{abu2010high, 10.1242/dev.202128} and single-particle tracking (SPT) \cite{mir2018dynamic} have been employed to estimate the fraction of Bcd molecules undergoing slow diffusion. However, this fraction varies depending on the definition of slow diffusion used in each study. In our work, we visualize clusters that are stable for at least one second and have an average size of approximately $0.4\:\mu$m, representing only a subset of the measured slowly diffusing molecules \cite{10.1242/dev.202128}. We likely observe only the ``slowest" fraction of moving particles (Fig.~\ref{figS3C}B), while the rest of the slow fraction might result from transient interactions with clusters or non-specific binding. Interestingly, fluorescence recovery after photobleaching (FRAP) and FCS studies \cite{10.1101/2024.01.30.578077} have found that only about $5\:\%$ of Bcd molecules constitute the immobile fraction.

A second albeit related fundamental question arising from our observations is how Bcd clusters establish and maintain a linear relationship to nuclear concentration. Addressing this requires examining the rate at which molecules approach the cluster boundary and how intranuclear diffusion parameters influence this rate. For stable clusters, the capture rate of molecules within the boundaries must balance the escape rate. The approach rate depends on the concentration-dependent diffusion properties of the molecules. However, the relationship between molecular escape rate from clusters and concentration remains unclear. Further exploration is needed to understand how concentration information is transmitted to the clusters.

$90-95\:\%$ of the Bcd protein in the nucleus is not in detectable clusters. Previous analysis \cite{10.1016/j.cell.2007.05.025} suggests that interpreting position based on this soluble fraction would be too slow to account for the observed dynamics and precision of transcription. We propose that the increased concentration and larger size of clusters facilitate the response of target genes to the Bcd gradient.

It remains unclear whether enhancers read concentration information directly from the clusters or merely serve as a medium for seeding clusters. If enhancers only seed clusters, the information content of the clusters could be interpreted directly or indirectly by the gene's promoter region. This implies that information transmission from the cluster to the promoter depends on their physical proximity \cite{10.1038/s41586-022-04680-7, 10.1038/s41588-018-0175-z}, making it an event limited by chromatin dynamics. Recent studies tracking clusters associated with transcriptional hubs have shown a correlation between cluster-promoter interaction and transcriptional burst enhancement  \cite{10.1016/j.cell.2023.12.005}. 

Whether cluster concentration affects the frequency or duration of such interactions remains an open question requiring careful quantitative studies. Early indications suggest no simple relationship exists \cite{10.1093/nar/gkad227}. Furthermore, a transcriptional microenvironment can be highly complex, with multiple enhancers interacting and communicating with the gene promoter via a single cluster \cite{10.7554/elife.45325}. How information stored in clusters is shared with DNA elements in the microenvironment remains an open question.   

Using a simple model \cite{10.1016/S0006-3495(77)85544-6} and our measured cluster properties, we demonstrate that clusters could potentially function as sensors at significantly faster timescales. This highlights a potentially crucial role for clustering in determining biological timescales, particularly for transcription. Specifically, clustering may offer an alternative explanation for noise suppression in molecular concentration readout. Currently, spatial and temporal averaging is the most commonly evoked scenario \cite{10.1016/j.cell.2007.05.025, 10.1103/physrevlett.103.258101}, which operates at considerably longer timescales. Clusters could provide a faster mechanism.

In conclusion, our analysis provides quantitative insights into cluster properties using fluorescently labeled living embryos, shedding light on the flow of biological information from the cell nucleus to a gene locus. This study emphasizes the potential role of clusters in maintaining nuclear concentration information and highlights their importance in the transcriptional response to morphogen gradients. While our work focuses on a specific transcription factor in a model organism, the principles we uncover are likely applicable across diverse biological systems, including mammals. 

Future research should explore the exact mechanisms by which clusters form and maintain their relationship to nuclear concentration. Investigating the dynamics of molecular interactions within clusters and their impact on transcriptional regulation will be crucial. Additionally, understanding the role of enhancers in seeding clusters and how cluster information is transmitted to promoters will provide deeper insights into gene regulation mechanisms. Advanced imaging techniques and quantitative models will be instrumental in addressing these questions, paving the way for broader applications and further research into the fundamental mechanisms of gene regulation.
%
\section*{Acknowledgments}
We thank Marianne Bauer, William Bialek, Clifford P. Brangwynne, Andrej Košmrlj, and Michal Levo for their discussions and comments on the manuscript. This work was supported in part by the U.S. National Science Foundation, through the Center for the Physics of Biological Function (PHY-1734030), and by National Institutes of Health Grants R01GM097275, U01DA047730, and U01DK127429, and the Howard Hughes Medical Institute.


\bibliography{ms}%

\onecolumngrid

\onecolumngrid

\clearpage
\newpage
\section*{\label{Meth}Materials and Methods}
\subsection*{Fly husbandry and genetics}
\textit{Drosophila} fly lines expressing \text{bcd-GFP} from \cite{10.1073/pnas.1220912110} were used as the starting point. In all such lines, the endogenous Bcd was replaced with a null phenotype \emph{Bcd\textsuperscript{\textnormal{E1}}}. Stable stocks expressing NLS-MCP-mRuby3 \: ; \: Bcd-eGFP-\emph{bcd \: \textsuperscript{\textnormal{E1}}} were created. Virgins from these stocks were then crossed with males expressing reporter constructs with the gene regulatory regions, while the gene body was substituted with MS2 stem-loop cassettes and LacZ. 

For the synthetic enhancers, the following scheme was used:
A 472 base pair (bp) fragment spanning the modified \textit{hb} proximal enhancer and the \textit{hb} P2 basal promoter was synthesized by IDT and ligated into the piB-hbP2-P2P-MS2-24x-lacZ-$\alpha$Tub3’UTR construct \cite{10.1016/j.cub.2013.08.054} between the restriction sites HindIII and NcoI. In the resulting reporter construct the hb promoter drives the expression of 24 copies of the MS2 loops and is followed by the lacZ coding sequence. The number of MS2 loops in the reporter was verified by Sanger sequencing. In the strong enhancer reporter, 8 suboptimal Bcd binding sites were converted to the consensus sequence TAATCC, resulting in a total of 11 strong Bcd binding sites. In the weak enhancer reporter, all 3 consensus sequence TAATCC were converted to the suboptimal Bcd binding site TAAGCT, resulting in a total of 11 weak Bcd binding sites. Both constructs were integrated into the 38F1 landing site on chromosome II of the fly line FC31 (y+); 38F1 (w+) using FC31 integrase-mediated cassette exchange \cite{10.1534/genetics.106.056945}. All fly lines from which males were crossed and their sources are tabulated in Table~\ref{table:1}.
%
%
\subsection*{Sample preparation} 
Embryos were harvested on apple juice plates, using protocols mentioned earlier \cite{10.1073/pnas.1220912110}. 
Staged two-hour-old embryos were dechorionated by hand by rolling them over a tape band (Scotch).
Dechorionated embryos were placed on the lateral side on a mounting membrane lined with glue. 
The glue was prepared by submerging $10\:$cm of Scotch tape in $4\:$ml Heptane for 48 hours in a shaker at $37^o$C.
A drop of glue was placed on the mounting membrane, gently smeared evenly, and was then allowed to air dry before placing the dechorionated embryos.
After the embryos were placed on the membrane, they were submerged in a mixture of halocarbon oil ($60\%$ Halocarbon27, $40\%$ Halocarbon700, Sigma), and then covered with a $25\times25\:$mm$^2$ glass coverslip (Corning). 
%
%
\subsection*{Imaging} Three-dimensional stacks of fluorescence images were acquired using the fast airyscan mode of a Zeiss LSM 880 microscope, run by Zen Black 2.3, SP1 software. 
A Plan-Apochromat 63x/1.4 oil immersion objective (Zeiss) was used for all measurements.
GFP was excited with the $488\:$nm line of the Argon laser ($140\:\mu$W), while mRuby3 was excited using the $561\:$nm diode-pumped solid-state laser ($36\:\mu$W).
Laser power at the back aperture of the objective was measured with a power meter  (PM100D, Thorlabs) at the beginning of each measurement session. 
The MBS 488/561 beamsplitter combined the beams.
The emission filter set, BP 420-480 / BP 495-550 was used for GFP emission, while  BP 495-550 / LP 570 was used for mRuby3.
The effective emission peak wavelengths were $515\:$nm for GFP and $578\:$nm for mRuby3.
A detector gain of 740 was used for all imaging cases.
The voxel size was fixed at $43 \times 43 \times 200 \: \mathrm{nm}^3$ for all 3D measurements.
For 2D single-plane videos, however, the z-section thickness was $1000 \:$nm.
The frame times were $497\:$ms for each frame for both color channels, with a pixel dwell time of $0.744 \:\mu$s.
Each image frame was $1044\times1044$ pixels, or $~45 \times 45 \:\mu$m for the 3D acquisitions. No averaging was done.
Imaging was done using the ``Fast Airyscan" mode,  with final images obtained after applying the "Airyscan Processing" within the Zen software.
\par
Imaging was conducted on embryos in nuclear cycle $\#14$, between the 20\textsuperscript{th} to the 35\textsuperscript{th} minute after mitosis. 
The nuclei at the embryo's surface facing the glass coverslip were imaged.
To ensure that the entire nucleus was scanned, a total z depth of $14\:{\mu}$m, with the central plane of the nucleus as the center was imaged. The stack was split into $70$ z-frames, with a $\sim500$ nm frame thickness. 
The horizontal dimensions of the images were $\sim 45 \times 45 \:\mu$m along the x-y plane, spanning $\sim40$ nuclei. Four such image stacks were recorded per embryo at various positions along the A-P axis.
%
%
\subsection*{Embryo fixation} Embryo fixation presented two challenges: 1) preserving the fluorescence of GFP after fixation, and 2) preserving the clusters themselves. To address both, we exclusively used freshly dissolved methanol-free formaldehyde (Thermo Scientific Pierce) at a final concentration of $4\:\%$ for embryo fixation. Throughout fixation and handling, we ensured that the embryos' exposure to organic solvents such as heptane, methanol, or ethanol was minimal. With these modifications to the standard protocol \cite{10.1371/journal.pbio.1000596}, fixation and visualization of Bcd clusters in the embryos can be achieved. 
%
\subsection*{Pixel correlation} To achieve pixel correlation, we separately autocorrelated the pixels along the x and y axes. We utilized the \texttt{crosscorr} function in \texttt{MATLAB} for this purpose. Although this function is typically used to determine the similarity between a time series and a lagged version of another series, in our case, we adapted it to find the autocorrelation of a pixel row (or column) with a lagged version of itself.
For pixel rows, (x), we get the correlation function ($c$) to be:
\begin{equation}\label{eq:corr}
c_{x,x} = 
\begin{cases}
  \frac{1}{T}\sum_{t=1}^{T-k} (x_n-\bar{x})(x_{n+k}-\bar{x}) & \text{k=0,1,2, ...} \\
  \frac{1}{T}\sum_{t=1}^{T-k} (x_n-\bar{x})(x_{n-k}-\bar{x}) & \text{k=0,-1,-2, ...}
\end{cases}
\end{equation} 
This is repeated over all the rows and the average is then calculated. 
The pixel columns (y) were similarly treated, after which the averages of the rows and columns were calculated. 
The correlation lengths calculated along the x-axis were equal to those calculated along the y-axis for all images.
The x- and y-axis data were then combined to obtain the overall image average.
The average function was fitted with an exponential, $y(x) = a+b\cdot exp(-c\cdot x)$ and the ``correlation length" was computed by $\lambda_{corr}=x_0+\frac{log(2)}{c}$.
Subsequently, the error in the correlation length is given by, $\sigma_{\lambda}=\lambda\cdot\frac{\sigma_c}{c}$.
This operation is selectively done for either the pixels exclusively within or outside the nuclear masks in the images.
%
%
\subsection*{Local maxima detection}
High-intensity foci of GFP-tagged proteins are scattered throughout the nucleus. Some of these foci result from protein clustering, while others are due to noise in the intensity. The centroids of these foci appear as local intensity maxima, and detecting them involves a two-step process. While the first step is applied only once, the second step is iteratively applied until the local maxima are located with high accuracy.
\par
In the first step, the nuclear pixels are segmented and an Otsu thresholding is performed. Only pixels with values above the threshold are retained, while the rest are converted to ``not a number" (NaN). 
The nuclear pixels are then rescaled to the interval $[0, 1]$, resulting in image $I^1$ (Fig.~\ref{figS1C}A, Top), to which the second step is applied.
\par
In the second step, local thresholding is applied. First, a $25\times25$ pixel window is created and the moving mean ($\mu_k$) and moving standard deviation ($\sigma_k$) are computed using the window on the nuclear pixels. 
This results in two matrices, one containing the moving means, $\mu_{I^1}$ (Fig.~\ref{figS1C}A, Middle row, left) and the other containing the moving standard deviations, $\sigma_{I^1}$ (Fig.~\ref{figS1C}A, Middle row, right), which are added ($\mu_{I^1} + \sigma_{I^1}$). This sum serves as the local threshold matrix, which is subtracted from $I^1$.
Pixels in $I^1$ with values below the corresponding cell in the local threshold matrix ($\mu_I^1 + \sigma_I^1$) are set to zero and the resulting image is rescaled to the interval $[0, 1]$ (Fig.~\ref{figS1C}A, Bottom row).
This gives the image, $I_2$, from which moving mean and moving standard deviation matrices are calculated and a new threshold matrix is generated ($\mu_{I^2} + \sigma_{I^2}$). This threshold matrix is subtracted from $I_2$ to obtain $I_3$ (Fig.~\ref{figS1C}B, Bottom row).
These steps are iteratively applied $m$ times yielding a set of images $I^1 \dotso I^m$.
With each local thresholding iteration, fewer pixels are retained around the local maximum, determining the center of the local maximum more accurately with each iteration. 
\par
To determine the optimal $m$ iterations required for optimal maxima localization, we first binarized the images $I^{\:i\in{[1,m]}}$ by setting all nonzero pixels to 1. 
In the resulting binarized images, $I_{bin}^{\:i\in{[1,m]}}$, we calculated the structural similarity index (SSIM) values using the built-in \texttt{MATLAB} function \texttt{ssim}, to assess the differences introduced in the images as a result of local thresholding.
Specifically, we computed pairwise SSIM values of the images $I_{bin}^i$ with respect to the image, $I_{bin}^1$ as the reference image.
The SSIM value drops with each $i\in{[1,m]}$, as the subsequent images are progressively poorly correlated with the starting image. 
However, at $m\sim15$, the SSIM values stabilized, indicating that further local thresholding would not improve the maxima detection.
\par
In the subsequent image, $I_{bin}^m$ was used to compute the location of the centroids of the local maxima.
This gave us the location of the intensity maxima in the nuclei with very high precision, although the maxima detected cannot be sorted by the size of the corresponding spots.
%
%
\subsection*{Pair-correlation} The local maxima in the nuclei are identified in all the frames of a video, and subsequently projected into a single map. The resulting time projection of the Bcd-GFP local intensity maxima has randomly dispersed points and focal accumulations of points in space. 
The randomly distributed points can be considered representative of a Poisson process and the focal accumulations can be modeled as Gaussian functions convolved with hypothetical singularities. 
To estimate the average density and effective size and the relative density of maxima within these focal accumulations representing a Gaussian process, we employ the pair-correlation function \cite{10.1371/journal.pone.0031457}.
\par
The density function of the points expressed in polar coordinates is given by $\rho(\vec{r})$.
The pair-correlation function for such a point distribution is given by \cite{10.1371/journal.pone.0031457}:
\begin{equation}\label{eq:paircorr1}
g(\vec{r}) = <\rho(\vec{R})\rho(\vec{R}-\vec{r}))>/\rho^2
\end{equation} 
Here, $\rho$ is the average density.
In practice, this correlation function is calculated using Fast Fourier Transforms applied to an image $I$ containing the point distribution:
\begin{equation}\label{eq:paircorr2}
g(\vec{r}) = \frac{1}{\rho^2}\cdot\frac{FFT^{-1}(\abs{FFT(I)}^2)}{ FFT^{-1}(\abs{FFT(W)}^2)}
\end{equation} 
Here, $I$ is a sparse matrix with $1\:$s at the locations of the maxima and $0\:$s elsewhere.
The quantity $W$ is a window matrix adjusted to fit within the area of a nuclear cross-section taken as a convex hull.
\par
If we consider the density of the Poisson process to be $1$, and the Gaussian process peak density to be $\rho ^{\prime}$ above $1$, we get the expression:
\begin{equation}\label{eq:paircorr3}
g(r) = \rho ^{\prime}exp(-r/\sigma ^2)+1
\end{equation} 
Here $\sigma$ denotes the size of the focal accumulation of the maxima, representing the Gaussian processes. 
This expression can then be used to fit the pair-correlation function to derive the effective width of the function and infer the increase in density within these Gaussian accumulations.
%
%
\subsection*{Positioning a nucleus in the embryo} 
To determine the position of a nucleus in the embryo we define a coordinate system. Initially, two images are acquired to get the full two-dimensional extent of the embryo: one of the embryo's anterior and one of the posterior halves, imaged at the midsagittal plane with otherwise identical imaging conditions as for the nuclei. From these two images we construct the compound image of the full embryo and identify (in software) the locations of the anterior $(x_0, y_0)$ and posterior $(x_L, y_L)$ tips of the embryo ($L$ is the length of the embryo) in microscope stage coordinates.
 \par
The line connecting these two points represents the anterior-posterior (AP) axis, or the x-axis in the embryo coordinate system ($X'$).
Perpendicular to this line is the y-axis ($Y'$), passing through $(x_0, y_0)$.
Hence, the anterior end is ($0,0$) and the posterior end is ($L,0$) in the embryo coordinates.
\par
Now, if the centroid of a nucleus is ($x_i, y_i$) in the microscope stage coordinates, the distance of the nucleus from $(x_0, y_0)$ is $r_i = \sqrt{(x_i-x_0)^2+(y_i-y_0)^2}$ and the angle made by $r_i$ with the AP axis is given by $\theta_i = arctan{(y_i-y_0)/(x_i-x_0)})-arctan({(y_L-y_0)/(x_L-x_0)})$.
Hence, the location of a nucleus in the $X'$ coordinates is given by $x_i' = r_i\cdot cos(\theta_i)$, or:
\begin{equation}\label{eq:locAP}
x_i' = \sqrt{(x_i-x_0)^2+(y_i-y_0)^2}\cdot cos\left(tan^{-1}\left(\frac{y_i-y_0}{x_i-x_0}\right)-tan^{-1}\left(\frac{y_L-y_0}{x_L-x_0}\right)\right)
\end{equation} 
This value can be computed for each nuclear centroid.
To pool nuclei by their position, nuclei with $x_i'$ within the position bin edges are accumulated.
%
%
\subsection*{Segmentation of a ``filled" nucleus} 
Nuclei segmentation is based on the GFP signal in Bcd-GFP-expressing embryos, which reduces the number of segmentable nuclei to those that are GFP-enriched, or ``filled", such that the nuclear boundary can be accurately identified utilizing the higher intensity of GFP within the nucleus. Automated segmentation of Bcd-GFP expressing nuclei was done using the following scheme:
First, the raw images were contrast adjusted using \texttt{imadjustn}, then filtered with a median filter, \texttt{medfilt3}, followed by a Gaussian filter, \texttt{imgaussfilt3}. 
A cuboidal structural element was then used for a series of morphological transformations to the resulting images. 
Erosion was applied (\texttt{imerode}), followed by a reconstruction, (\texttt{imreconstruct}), and then dilation, (\texttt{imdilate}).
The complement of the reconstructed image was obtained (\texttt{imcomplement}) and then blurred with a Gaussian filter (\texttt{imgaussfilt3}).
The resulting image was closed (\texttt{imclose}) and eroded, and a binary mask for the nucleus pixels was subsequently obtained. 
Finally, watershed segmentation (\texttt{watershed}) was applied to separate any conjoined neighboring nuclei. Labels were then assigned to the nuclear masks to identify individual nuclei.
\par
\vspace{.2 cm}
%
%
\subsection*{Locating clusters (in 3D)} 
A technique for 2D local maxima localization was introduced in a previous section entitled \textit{Local maxima detection}. 
This technique indiscriminately detects all local intensity peaks, including noise spikes and protein clusters.
The difference between a ``real" cluster and a ``noise-related local maximum" is that the spot size for a cluster is at least as large as the point spread function, while a noise-related maximum is likely to be smaller. 
To capture this difference, we chose a pixel size smaller than the Nyquist criterion of a diffraction-limited spot. 
With an x-y size of $43\:nm$, a diffraction-limited spot spans $4-5$ pixels in either direction.
Hence, a cutoff limit of 3 pixels effectively differentiates a cluster from a noise-related maximum. 
For z-slices, we chose a thickness of $200\: nm$. 
Since the PSF width along the z direction is larger than $500\: nm$, any maxima that does not span at least 2 z-slices are likely noise-related maxima. 
Hence we chose 2 pixels as the cutoff limit along the z axis.
\par
For live imaging of mobile structures like subdiffusive clusters, the likelihood of detecting a cluster in two consecutive frames depends on the frame rate. As shown in Fig.~\ref{figS1D}H, the detection probability of a cluster in frames imaged $\sim500 \:ms$ apart is greater than $70\:\%$.
Therefore, for z-slices imaged $200\:ms$ apart with thickness less than the PSF width ($\sim 500$ nm), we should be able to detect the cluster with high reliability.
\par
To identify only relevant puncta-like entities in the nucleus, we employed the following technique on the raw images of Bcd-GFP nuclei.
First, morphological top-hat filtering was applied using a ``disk" as the structural element to the 3D raw images of the nuclei using top-hat filtering. 
The transformed image thus obtained was used to detect local intensity maxima peaks.
For this, the top 1 percentile pixels within a nucleus were selected from the transformed images. 
Joined neighboring spots were then separated by applying a watershed algorithm.
\par
It can be argued that the centroid of the local maxima peaks from the raw images is preserved through this morphological transformation. 
Next, the spot mask is obtained from the spot segmentation in the morphologically transformed image, and the mask is then applied to the raw image.
Intensity weighted centroids in 3D of the voxels within each mask are then calculated using \texttt{WeightedCentroid} on the raw image.
This gives the peak position of each cluster. 
\par
However, not all clusters thus detected are retained for further analysis. 
A size thresholding (as mentioned above) is then performed such that if the x-y cross section of a detected spot is less than $3\times3$ pixels wide and the z depth is not at least 2 pixels wide, the spot is discarded.
That brings the threshold volume to $3\times3\times2=18$ pixels.
A corresponding effective spot diameter $d$ can be calculated from the threshold volume, such that $d = ({6}/{\pi}\times vol)^{1/3}$. 
The threshold diameter turns out to be $3.25$ pixels wide, which converted to absolute units gives, $138.2\:nm$, which is significantly less than the 3D PSF of the microscope.
Thus, using this technique we identify the locations (only) of the ``real" puncta in the Bcd-GFP nuclei.
\vspace{.2 cm}
\par
%
%
\subsection*{Fitting clusters} 
To extract cluster-relevant parameters cluster fitting is performed on the raw image pixels. 
Although a cluster is a three-dimensional entity spanning multiple imaging sections, we perform a two-dimensional fitting of the intensity profile in the plane passing through the intensity-weighted center along the z-axis. 
Given that the resolution along the z-axis is approximately five times poorer than along the x-y axes, any fitting along the z-axis introduces significantly higher errors.
Pixels within a square window centered on the cluster centroid are chosen from the plane of the cluster centroid's z-coordinate. 
The window length is set at $2w+1$ pixels, where $w$ is $\sim 12$ pixels for a typical window size of $1.1.\times1.1\:\mu m^2$.
\par
To fit the intensity profile within this window, a 2-dimensional Gaussian function is employed \cite{10.1038/nmeth.3446}. The fitting procedure utilizes least-square curve fitting (\texttt{lsqcurvefit}) with the \texttt{levenberg-marquardt} algorithm. 
The Gaussian fitting equation used is:
\begin{equation}\label{gauss1}
\begin{split}
f(x,y) = I_a exp(-(a(x-x_0)^2+ \\ 2b(x-x_0)(y-y_0)+c(y-y_0)^2))+I_{bg}
\end{split}
\end{equation}
where, 
\begin{equation}\label{gauss1Para}
    \begin{split}
    a = \frac{cos^2\theta}{2\sigma_1^2} + \frac{sin^2\theta}{2\sigma_2^2} \\
    b = \frac{sin2\theta}{4\sigma_1^2} + \frac{sin2\theta}{4\sigma_2^2} \\
    c = \frac{sin^2\theta}{2\sigma_1^2} + \frac{cos^2\theta}{2\sigma_2^2}
    \end{split}
\end{equation}

Here, $I_a$  represents the intensity amplitude, and $I_{bg}$ is the background intensity level.
The initial guess value for $I_a$ was the pixel value at the center of the window, and $I_{bg}$ was approximated as $I_{nuc}$, the average intensity of the nucleus. 
The initial guesses for both, $\sigma_1$ and $\sigma_2$ are $w/2$, and the rotational angle $\theta$ is initialized to $0$.
Bounds for $\sigma_1$ and $\sigma_2$ are set to $w^2$ and the bounds to $x_0, y_0$ are set to $-w:w$.
The angle $\theta$ is constrained between $0<\theta<\pi/4$.
Candidates whose fitted parameters do not meet the criteria are automatically discarded as cluster candidates ($<2\:\%$).

It's important to note that $I_a$ and $I_{bg}$ are obtained separately from the fits, ensuring that $I_a$ is automatically background-corrected. 
%
%
\subsection*{Cluster properties} 
Using the parameters obtained from the fits, we obtain the measures for three cluster properties: cluster size, $d$, the concentration of Bcd molecules within a cluster, and the total molecules inside a cluster. For the notations used in this section, refer to the parameters obtained in the previous section.
\par
To calculate the effective size of clusters, we consider the Gaussian spread along x and y directions ($\sigma_x, \sigma_y$).
The effective radius, $r_{eff}$ is derived as, $r_{eff} = \sqrt{\sigma_1^2 + \sigma_2^2}$. This represents the effective size of the three-dimensional cluster, assuming the Gaussian width as a projection on a section of the obloid-shaped cluster representing the actual spot.
\par
The intensity amplitude $I_a$ of the cluster gives an estimate for the concentration of Bcd molecules in the cluster.
To obtain an estimate for the total number of molecules within a cluster we consider the quantity, $I_c = 2 \pi I_a \sigma_1 \sigma_2$. 

After obtaining the measures for the cluster properties, compute their nuclear averages, $\expect{d}, \expect{I_a}, \expect{I_{bg}}$, and $\expect{I_c}$. These nuclear averages of the measures of cluster properties are then examined for correlation with the nuclear Bcd concentration, given by $I_{nuc}$, and the position of the nucleus in the embryo $x/L$.
Subsequently, the nuclear average cluster property values are discretized into equidistant bins of either the normalized nuclear positions ($x/L$) or the average nuclear concentration ($I_{nuc}$). For each bin, an equal number of bootstrap data samples are drawn and the mean and the standard deviations are separately computed.

%
%
\subsection*{Slope calculation} 
The nuclear averages of the measures of cluster properties ($\expect{d}, \expect{I_a}, \expect{I_{bg}}$ and $\expect{I_c}$) (or their natural logarithm) are plotted against the corresponding nuclear Bcd concentration, $I_{nuc}$ (or the nuclear position, $x/L$).Linear regression models are fitted to the data, and parameters such as the coefficient of determination $R^2$, the slope of the linear fit, and the error in the slope are derived from these models.
\par
The nuclear averages of the measures of cluster properties are assumed to linearly correlate with $I_{nuc}$, whereas their dependence on $x/L$ follows an exponential pattern. Therefore, the slope of the linear fit of the natural logarithm of the nuclear averages of the measures of cluster properties, plotted against $x/L$ can be utilized to determine the exponential decay constant $\lambda$, such that 
$\lambda = -(1/slope) \pm \sigma_{slope}/slope^2$.
%
\subsection*{Error in nuclear property estimation using cluster property}
To estimate the error in nuclear concentration estimation $\sigma_c$ using cluster properties like $\expect{I_a}$, $\expect{d}$, and $\expect{I_c}$, we use the slope ($s$) obtained from the linear fits and the error in the cluster property estimation in each concentration bin $\sigma_i$.
The formula for $\sigma_c$ is derived as $\sigma_c = \sigma_i/s$. Additionally, the uncertainty associated with each with each $\sigma_c$ ($\sigma_{\sigma_c}$) is $\sigma_{\sigma_c}=\sigma_c\cdot\sqrt{(\sigma_{\sigma_i}/\sigma_i)^2+(\sigma_s/s)^2}$, where, $\sigma_{\sigma_i}$ represents the error associated with cluster property determination, and $\sigma_s$ denotes the error in the slope ($s$).
Similarly, the error in estimating nuclear position ($\sigma_p$) using cluster properties is determined using the exponential decay constant ($\lambda$) as $\sigma_p=\lambda\cdot\sigma_i/i$, where $\lambda$ represents the exponential decay constant and $i$ denotes the average nuclear position.  The uncertainty in $\sigma_p$ is given by $\sigma_{\sigma_p}=\sigma_p\cdot\sqrt{(\sigma_{\sigma_i}/\sigma_i)^2+(\sigma_i/i)^2+(\sigma_\lambda/\lambda)^2}$.
All these expressions are derived from the laws of error propagation.
%
%
%

\subsection*{Segmentation of nuclei using MCP-mRuby3 intensity} 
While Bcd-GFP enriches the nucleus, the relative intensity of MCP-mRuby3 is higher in the cytoplasm than in the nucleoplasm, making the nuclei appear ``hollow".
Segmentation of MCP-mRuby3 expressing nuclei involved an approach distinct from Bcd-GFP expressing nuclei. Since MCP-mRuby3 nuclei lack fluorophores, their signal is lower than that of the internuclear space. To segment these nuclei, we first adjusted the image brightness using \texttt{imadjustn}, which increases image contrast by adjusting intensity values. Next, we applied a three-dimensional median filter followed by a Gaussian filter. An extended regional maxima transformation (\texttt{imextendedmax}) was then performed. The resulting image was binarized, and the inverse of the binary image was created. Subsequently, a three-dimensional kernel was convolved with the binary image, and a threshold was applied. Watershed segmentation was applied to the thresholded image. The resultant image underwent opening using a cuboidal structural element, and any holes within the bright structures were filled to generate the final mask for the nuclei.
\par 
To match a Bcd-GFP expressing ``hollow" nucleus with an MCP-mRuby3 expressing  ``filled" nucleus, we checked if their centroids were within half a nuclear length of each other. Nuclei that couldn't be mapped in this manner were discarded from further analysis.
\par
\vspace{.2 cm}
\subsection*{Transcription hotspot detection} Transcription hotspots are nascent mRNA accumulations at the site of active transcription in the nucleus. These nascent mRNA molecules have MS2 stem-loops that are bound by MCP-mRuby3 fusion proteins. The mRuby3 fluorescence intensity lets hotspots appear as bright spots within the nucleus against a darker background. To identify these spots, the nuclear boundaries were determined using a hollow nuclear segmentation approach (see above). Within these boundaries, the identification of transcription hotspots began with the application of a Difference of Gaussian (DoG) algorithm to the raw images. The resulting image was convolved with the original raw image and then rescaled. A threshold based on the nuclear pixel intensity ($I_{nuc}$) was applied, discarding any pixels with intensities lower than $4\sigma_{I_{nuc}} + \mu_{I_{nuc}}$. Subsequently, masks representing potential transcription hotspots were generated, and a size cutoff of $18$ pixels was imposed on these masks.
%
%
\subsection*{Radial intensity profile and coupling fraction calculation} 
To calculate the distance limits for a Bcd cluster to be coupled with an mRNA hotspot, we first calculate the width of the radial intensity profile of Bcd-GFP around the mRNA hotspot. Bcd-GFP clusters associated with a transcriptional hotspot might coincide with the hotspot, in which case the distance is given by the centroid distances of the two fluorescence accumulations. 
\par
Firstly, the mRNA hotspots were segmented, and their intensity-weighted centroids were determined by using the \texttt{regionprops} function of \texttt{MATLAB}.
Next, the intensity of Bcd-GFP ($I_r$) was computed, $r$ being the radial distance from the transcription hotspot centroid.
The intensity profile was computed exclusively on the x-y plane passing through the hotspot centroid.
$I_r$ was obtained by averaging pixel intensities within a ring from $r$ to $r+0.1\: \mu m$.
Data for each $I_r$ was aggregated across multiple nuclei from various embryos to generate an average $I_r$ profile (Fig.~\ref{figS4B}B, C (cyan, error bars)).
\par
To determine the accumulation radius for a gene, the average $I_r$ profile was fitted with a double Gaussian function (Fig.~\ref{figS4B} C (cyan, solid line)):
\begin{equation}\label{gauss2}
f(x) = k_0 e^{\left(-{\frac{(x-x_0)}{a_0}}\right)^2} + k_1 e^{\left(-{\frac{(x-x_1)}{a_1}}\right)^2}
\end{equation}
Here, $r_0$ the accumulation radius, was defined as the full width at half maximum (FWHM) of the first Gaussian component. The error in $r_0$ was calculated from the fitting error.

Furthermore, the distance from the intensity-weighted centroid of the nearest Bcd-GFP cluster to the transcription hotspot centroid was measured. A histogram of these distances was plotted, and the cumulative probability function was derived directly from the histogram or through spline fitting. The cumulative probability value at $x=r_0$ provided the coupling fraction for a gene (Fig.~\ref{fig4}H).
\par
Alternatively, the histogram of the nearest neighbor TF cluster distances from the mRNA hotspot could be fitted with double Gaussians (Fig.~\ref{figS4B} C (purple, broken lines)). The first peak corresponded to the nearest neighbor cluster coupled to the gene, and the second, weaker peak indicated the cluster nearest to the gene that was not coupled. The intersection of these two Gaussian fits marked the boundary $r_0^`$ (Fig.~\ref{figS4B} C (black, broken lines)), ensuring only clusters inside this boundary were considered coupled.
\par
The radius $r_0$ is influenced by several factors that contribute to broadening. Chromatin, not being stationary but subdiffusive in the nuclear space, causes motion blurring of point sources during video capture. Additionally, mRNA hotspots, consisting of multiple MS2 stem loops spread across the gene body, can span several kilobases at any given moment. Moreover, the stem-loops project out of the gene body with their own degrees of freedom. Transcription, being a kinetic process, allows stem loops to traverse linearly along the gene body at the speed of transcription (approximately $2 kB/min$). These factors collectively broaden the MS2 hotspot signal, necessitating its approximation as a point source convolved with a Gaussian to incorporate all forms of broadening.
%
%
\subsection*{Estimation of molecules per cluster and the total cluster fraction}
In a previous study, the total number of Bcd molecules in the nucleus was estimated using Western blots \cite{10.7554/elife.28275}. 
However, the construct was such that the Bcd concentration was uniform throughout the embryo, unlike the exponential decay observed along the axis in the wild-type gradient.
Given that the total number of Bcd molecules in the flat expression lines remains equivalent to the total expressing molecules in the wild-type gradient, we can establish a relationship between these quantities as follows:
\begin{equation}\label{integration}
N_{0_{flat}} \int_{0}^{L}\,dx\ = N_{0} \int_{0}^{L} e^{-x/\lambda}\,dx\
\end{equation}
Here, $N_{0_{flat}}$ represents the average concentration of Bcd in a nucleus in the flat expression lines used in the study, while $N_{0}$ denotes the concentration at $x=0$ for the wild-type line, which exhibits an exponential gradient with a length constant $\lambda$.
\par
Solving this equation yields $N_{0}=N_{0_{flat}}L/\lambda$. 
Using approximate values $N_{0_{flat}}\sim8000$, and $\lambda\sim0.2$, we obtain $N_0\equiv40000$ molecules. Assuming an average nuclear diameter of $5\:\mu m$, the average density of Bcd molecules in the nucleus is approximately $600$ molecules$/\mu m^3$.
\par
Given that an average cluster has a concentration of Bcd that is $2.2$ times higher than the nucleoplasm, the Bcd concentration inside a cluster at the anterior of the embryo is $1320$ molecules$/\mu m^3$.
With an average cluster diameter of $\sim 0.4 \:\mu m$, the average volume is $0.03\:\mu m^3$. 
By knowing the molecular density within a cluster and its volume, we calculate that there are approximately $37$ molecules per cluster at the anterior of the embryo.
An estimate of the molecules per embryo along the embryo axis is shown in Fig.~\ref{figS3C}C.
%
%
\subsection*{Derivation of the concentration sensing limit}
We consider a spherical cluster with an effective diameter $d$, containing a concentration of Bcd molecules $c_{clust}$, and Bcd's diffusion constant represented by $D$.
The fractional error $\delta N$ in counting $N$ molecules within the cluster is given by
\begin{equation}\label{eq:1}
\frac{\partial N}{N} = \left(\frac{6}{D\cdot d \cdot c_{clust} \cdot T_{clust}}\right)^{1/2}
\end{equation}
Rearranging EQ. \ref{eq:1} we get
\begin{equation}\label{eq:2}
T_{clust} = \frac{6}{D\cdot d \cdot c_{clust}} \left(\frac{\partial N}{N}\right)^{-2}
\end{equation}
The time $T_{clust}$ in EQs. \ref{eq:1} and \ref{eq:2} represents the duration required for a cluster to ``measure" the nuclear concentration with an accuracy of $\left(\frac{\partial N}{N}\right)$.
\par
The analogous time for a binding site of length $a$ to measure the nuclear concentration is given by:
\begin{equation}\label{eq:3}
T_{en} = \frac{1}{D\cdot a \cdot c_{nuc}} \left(\frac{\partial N}{N}\right)^{-2}
\end{equation}
Here, $c_{nuc}$ is the nuclear concentration.
Using EQs. \ref{eq:2} and \ref{eq:3} we derive the ratio:
\begin{equation}
    \frac{T_{en}}{T_{clust}}=\frac{d}{6a}\left(\frac{c_{clust}}{c_{nuc}}\right)
\end{equation}
The length of a typical binding site can be considered as $a=3.4\:nm$ \cite{10.1016/j.cell.2007.05.025}.
However, $d$, $c_{nuc}$, and $c_{clust}$ depend on the nuclear position, thereby making ${T_{en}}/{T_{clust}}$ a position-dependent quantity.
%
%
\subsection*{Point spread function measurements.} To determine the Point Spread Function (PSF) of the imaging system, $100\:nm$ fluorescent polystyrene beads (Thermo Fisher Catalog T14792) were imaged using the fast \textit{Airyscan} mode on a Zeiss LSM 880 microscope. The objective is a 63x objective (Zeiss Plan-Apochromat 63x-1.4 oil immersion). Beads were illuminated with a $488\:nm$ argon laser line. Images were acquired with a voxel size of $42\times42\times42\:nm$. A total thickness of $1\:\mu m$ was imaged with the beads at the center. The beads are mounted on a flat glass surface and hence all reside on the same imaging plane. Each field of view contains 20--30 beads. Several such images were acquired. Pixel correlation was performed on these images, and the average correlation length provided the point-spread function of the system.

\subsection*{Sequences for strong and weak enhancer constructs}

WT variant

\begin{quote}
  \ttfamily
  \fixsplit{80}{CACGCTAGCTGCCTACTCCTGCTGTCGACTCCTGACCAACGTAATCCCCATAGAAAACCGGTGGAAAATTCGCAGCTCGCTGCTAAGCTGGCCATCCGCTAAGCTCCCGGATCATCCAAATCCAAGTGCGCATAATTTTTTGTTTCTGCTCTAATCCAGAATGGATCAAGAGCGCAATCCTCAATCCGCGATCCGTGATCCTCGATTCCCGACCGATCCGCGACCTGTACCTGACTTCCCGTCACCTCTGCCCATCTAATCCCTTGACGCGTGCATCCGTCTACCTGAGCGATATATAAACTAATGCCTGTTGCAATTGTTCAGTCAGTCACGAGTTTGTTACCACTGCGACAACACAACAGAAGCAGCACCAATAATATACTTGCAAATCCTTACGAAAATCCCGACAAATTTGGAATATACTTCGATACAATCGCAATCATACGCACTGAGCGGCCACGAAACGGTAGGA}
  \end{quote}

All weak variant

\begin{quote}
  \ttfamily
  \fixsplit{80}
{CACGCTAGCTGCCTACTCCTGCTGTCGACTCCTGACCAACGTAAGCTCCATAGAAAACCGGTGGAAAATTCGCAGCTCGCTGCTAAGCTGGCCATCCGCTAAGCTCCCGGATCATCCAAATCCAAGTGCGCATAATTTTTTGTTTCTGCTCTAAGCTAGAATGGATCAAGAGCGCAATCCTCAATCCGCGATCCGTGATCCTCGATTCCCGACCGATCCGCGACCTGTACCTGACTTCCCGTCACCTCTGCCCATCTAAGCTCTTGACGCGTGCATCCGTCTACCTGAGCGATATATAAACTAATGCCTGTTGCAATTGTTCAGTCAGTCACGAGTTTGTTACCACTGCGACAACACAACAGAAGCAGCACCAATAATATACTTGCAAATCCTTACGAAAATCCCGACAAATTTGGAATATACTTCGATACAATCGCAATCATACGCACTGAGCGGCCACGAAACGGTAGGA}
\end{quote}

All strong variant

\begin{quote}
  \ttfamily
  \fixsplit{80}
{CACGCTAGCTGCCTACTCCTGCTGTCGACTCCTGACCAACGTAATCCCCATAGAAAACCGGTGGAAAATTCGCAGCTCGCTGCTAATCCGGCCATCCGCTAATCCCCCGGATAATCCTAATCCAAGTGCGCATAATTTTTTGTTTCTGCTCTAATCCAGAATGGATTAAGAGCGTAATCCTTAATCCGCGATCCGTAATCCTCGATTCCCGACCGATCCGCGACCTGTACCTGACTTCCCGTCACCTCTGCCCATCTAATCCCTTGACGCGTGCATCCGTCTACCTGAGCGATATATAAACTAATGCCTGTTGCAATTGTTCAGTCAGTCACGAGTTTGTTACCACTGCGACAACACAACAGAAGCAGCACCAATAATATACTTGCAAATCCTTACGAAAATCCCGACAAATTTGGAATATACTTCGATACAATCGCAATCATACGCACTGAGCGGCCACGAAACGGTAGGA}
\end{quote}

\clearpage
\onecolumngrid

\section{Supplemental Figures}
\renewcommand\thefigure{S\arabic{figure}} 
\renewcommand\theequation{S\arabic{equation}}
\setcounter{figure}{0} 

\begin{figure*}[h!]
\centering

\includegraphics[width=0.7\paperwidth]{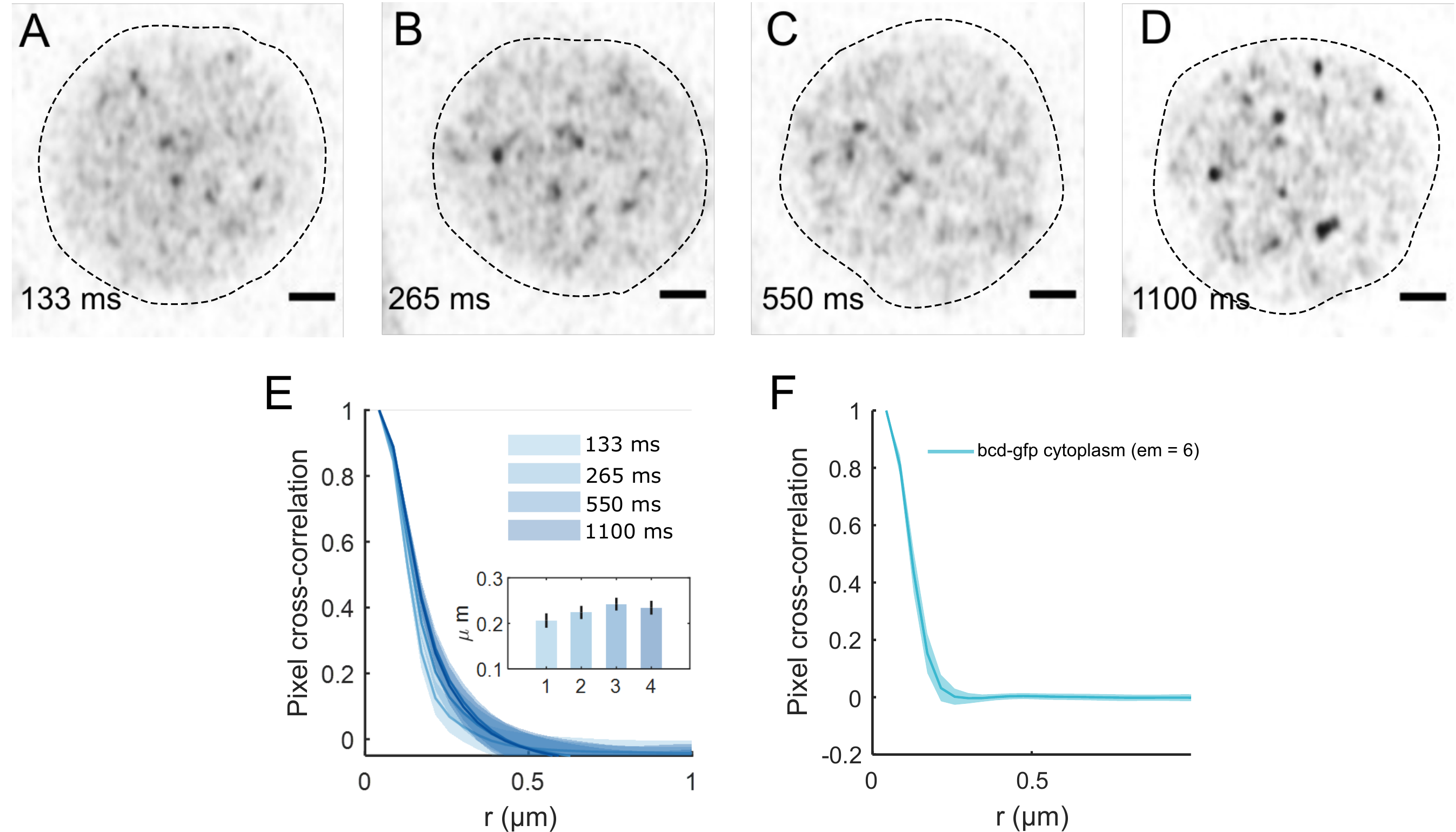}
\caption{\textbf{Imaging times and correlation lengths.} 
(A-D) Cross-sections passing through the centers of representative individual Bcd-GFP expressing nuclei, imaged at varying imaging times per frame (see annotation). Confocal \textit{Airyscan} microscopy was used in the \textit{fast} setting. Longer times per frame result in augmented mobile features blurring into the background, leaving only the brighter and more stable features visible (see increased contrast with $1100\:$ms setting). Dotted lines serve as guides to the eye for nuclear boundaries, and scale bars are $1\:\mu$m. 
(E) Pixel correlations are plotted for various times per frame, with the mean values and the standard errors shown as error bars in the inset. The correlation lengths are $0.21\pm0.02\:\mu m$, $0.22\pm0.02\:\mu m$, $0.24\pm0.02\:\mu m$ and $0.23\pm0.02\:\mu m$ for $133$ ms, $265$ ms, $550$ ms and $1100$ ms frame rates respectively. The correlation lengths are statistically indistinguishable, suggesting that motion-induced blurring does not significantly broaden resolvable structures, even at a time per frame of $\sim 1\:$s.
(F) Pixel correlation of cytoplasmic Bcd-GFP. We measure a correlation length of $0.19\pm0.02\:\mu$m, a value comparable to the width of the PSF, indicating that cytoplasmic Bcd mostly undergoes diffusive motion without forming clusters. 
}
\label{figS1A}
\end{figure*}

\begin{figure*}[h!]
\centering
\includegraphics[width=0.5\paperwidth]{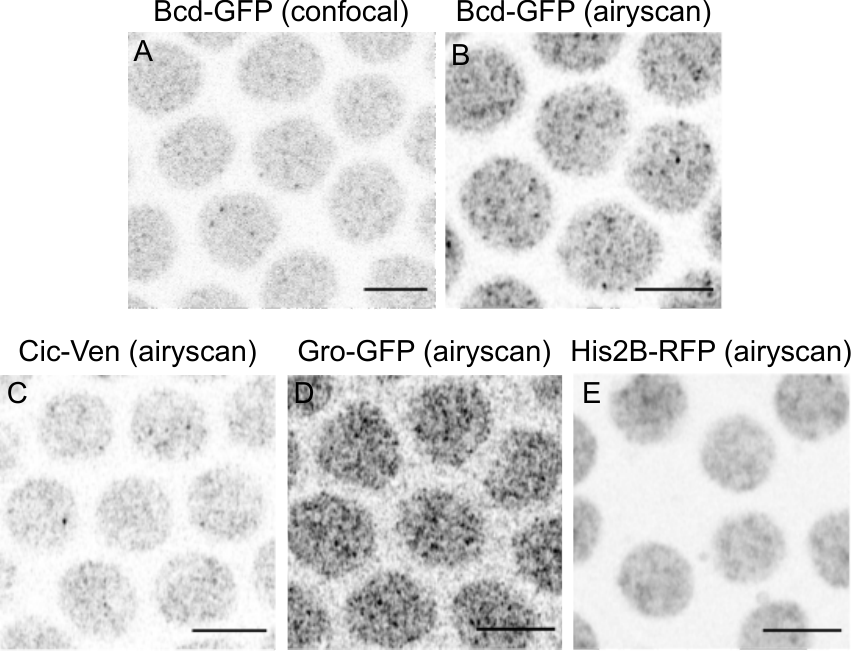}
\caption{\textbf{Various live protein labeling in \textit{Drosophila} nuclei.} 
(A-B) Traditional confocal (A) as well as \textit{Airyscan} confocal  (B) images (Zeiss) showing cross sections of Bcd-GFP expressing nuclei in living \textit{Drosophila} embryos during nuclear cycle 14. The pixel dwell times were the same for both images. Visual inspection identifies a higher signal-to-background ratio for cluster-like features in the \textit{Airyscan} image than with regular confocal microscopy. This is an indication that the higher resolution of the \textit{Airyscan} mode resolves smaller structures like the clusters better.
(C-E) \textit{Airyscan} confocal images showing nuclei expressing Capicua tagged with Venus, Groucho tagged with monomeric eGFP, and Histone2B tagged with RFP (otherwise imaging conditions as above). All scale bars are $5\:\mu$m.
}
\label{figS1B}
\end{figure*}

\begin{figure*}[h!]
\centering
\includegraphics[width=0.8\paperwidth]{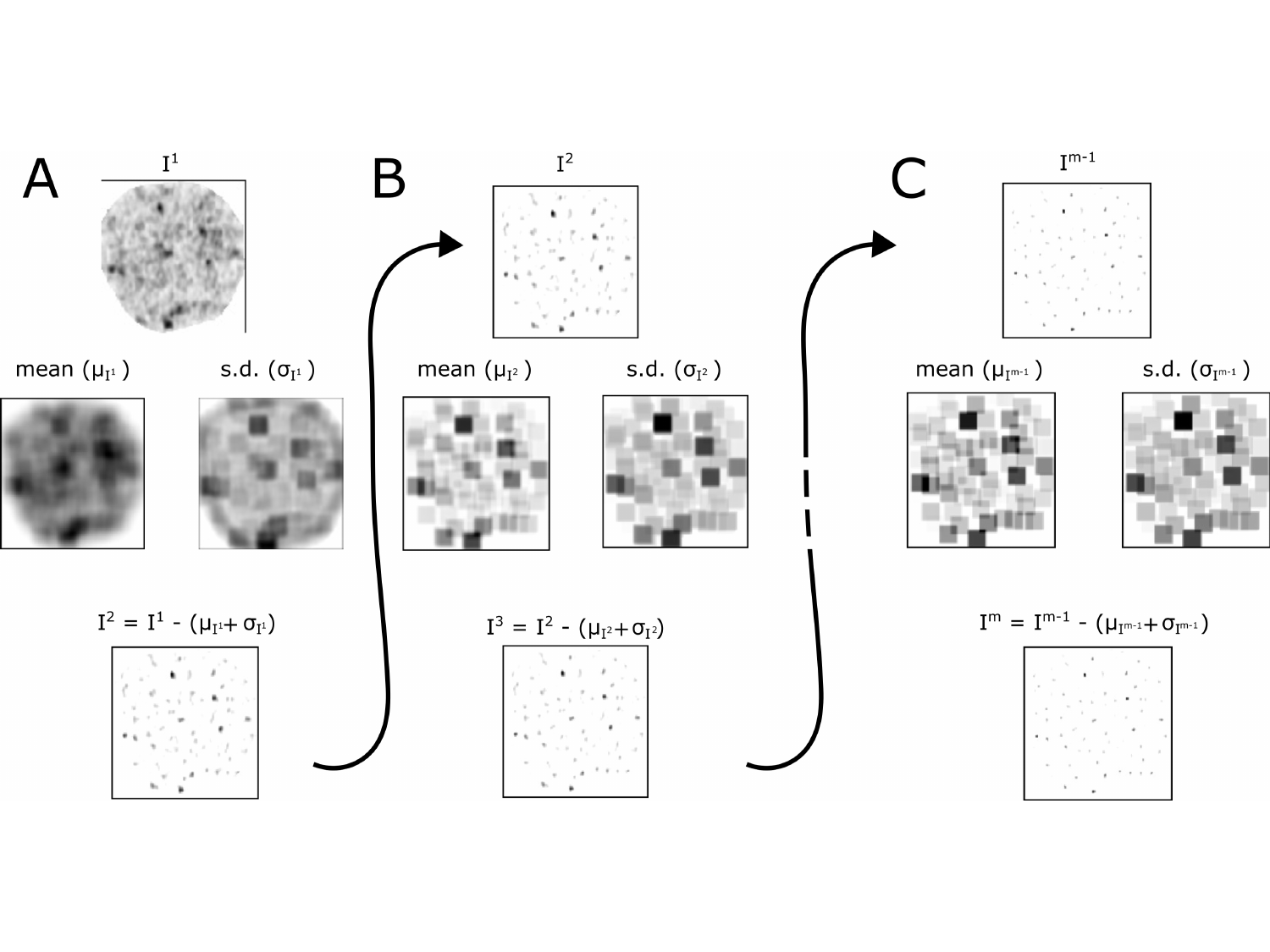}
\caption{\textbf{Local maxima detection algorithm.} 
(A) The top image displays the image of a nucleus expressing Bcd-GFP after Otsu thresholding. In the middle panel, the left image shows the moving means of the image at the top, computed over  $25 \times 25$ pixel windows.  The right image of the panel shows the moving standard deviation, over windows of $25 \times 25$ pixels computed on the image above. The bottom image is obtained by adding the middle row images and subtracting the resulting matrix from the top image. This resultant image is used for the subsequent operations in B. 
(B) The panel illustrates the recursive application of the process explained in A. This yields the image that acts as the input for the next layer of operations.
(C) Finally, the image obtained after $m$ successive application of local thresholding (Bottom) is utilized to locate the centroids of the local maxima.
}
\label{figS1C}
\end{figure*}
\begin{figure}[b!]
\centering
\includegraphics[width=0.8\paperwidth]{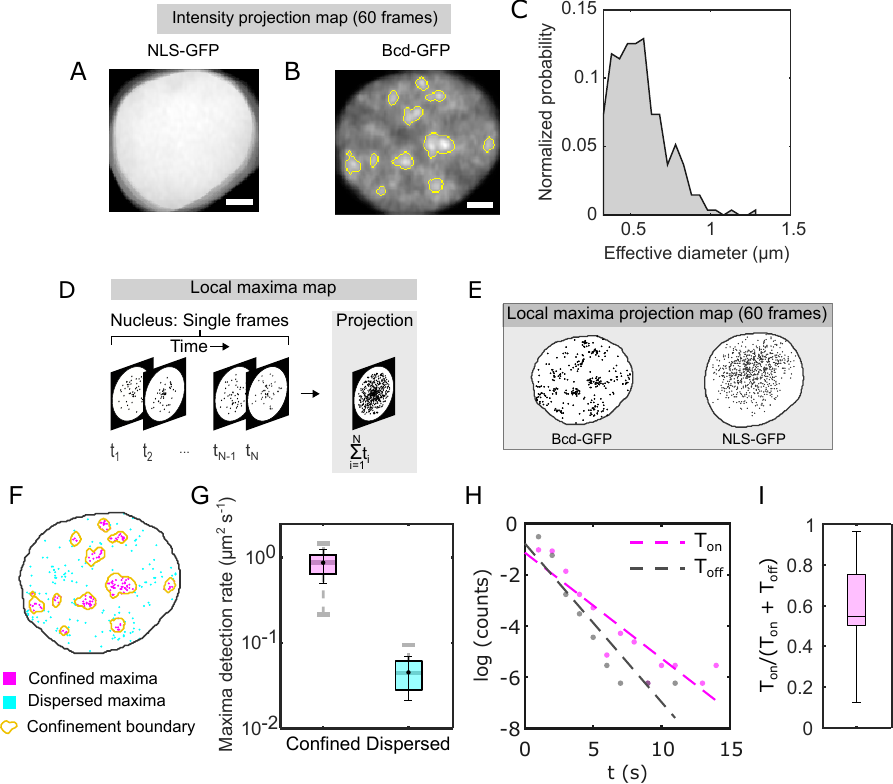}
\caption{\textbf{Frequency of cluster formation and lifetime of clusters.} (A, B) Time-projected images of a cross-section of a single nucleus expressing NLS-GFP (A) or Bcd-GFP (B). Movies are composed of 60 video frames, two per second. 
The nuclear pixels are segmented, and the centroid of the binary nuclear mask is used to register the nucleus so that the centroids across all frames are aligned. These registered nuclei from each frame are then projected onto a single frame to obtain the images seen in A and B.
The uniformity of the nuclear pixels in A image suggests that any heterogeneity in individual frames is random and thus averaged out over time. In contrast, the B image shows distinct subnuclear accumulations. These accumulations indicate that local intensity maxima appear over multiple frames in those regions. These accumulations are segmented (Materials and Methods), and their boundaries are shown as yellow line overlays. These boundaries represent confinement areas where local maxima frequently appear in the movies. The scale bar is $1\:\mu$m. 
(C) To calculate the effective sizes of the confinement areas marked in yellow in B, the area of each detected area in pixels is calculated. An effective diameter of each area, as an approximated circle is calculated. A histogram of the respective diameters is shown in C, with $mean\pm s.d.$ of $0.53\pm0.16\:\mu$m.
(D) Schematic showing a map of local fluorescence intensity maxima inside a nucleus (left). The local maxima maps are extracted from individual frames of $\sim30\:$s long videos (60 frames) of nuclear cross sections ($1\mu$m thick). All maps from a given video are projected onto a single frame to form the local maxima map (right). See also Materials and Methods. 
(E) Representative local maxima maps for a Bcd-GFP nucleus (left) and an NLS-GFP nucleus (right) extracted from nuclei imaged for 30 seconds (60 video frames). [Captions continued on the next page]
}
\label{figS1D}
\end{figure}

\addtocounter{figure}{-1}
\begin{figure} [t!]
\caption{
(F) The local intensity map of Bcd-GFP shown in E is repeated here. Utilizing the confinement area boundaries (in yellow), we can segregate the maxima into either confined (magenta) or dispersed (cyan) maxima. The magenta maxima occur with higher spatial density than the cyan ones. This is an indication that the confined maxima are more persistent and are candidates for clusters.
(G) This disparity in spatial density of maxima is quantified here. Box plots show the rate of detection of confined (magenta) and dispersed (cyan) maxima in the projection maps (44 nuclei, 12 embryos). Boxes extend from the 25\textsuperscript{th} to the 75\textsuperscript{th} percentile, while the horizontal divider marks the median. Mean and standard deviations are overlaid in black ($0.87\pm0.37$ and $0.04\pm0.02$ for confined and dispersed maxima respectively). These represent the spatial density of maxima of the full video projection. The spatial density of maxima per unit is obtained by dividing by the total time of the video (30 seconds), with unit $\mu m^2s^{-1}$. The maxima in cyan could be either noise or represent highly transient or mobile clusters. The high density of maxima in magenta indicates that a cluster is detected in multiple video frames within the yellow confines. 
(H) In this figure, we estimate the time intervals for which confined maxima are detected ($T_{on}$), as well as the time interval between two successive maxima detection ($T_{off}$). These can help understand the persistence of a confined maximum, and thereby a cluster. The magenta data shows a scatter plot of the natural logarithm of the probability of $T_{on}$ obtained by binning the $T_{on}$ data into equidistant bins. An exponential fit to the data (magenta broken line) gives the time constant associated with $T_{on}$, $\tau_{on} = 2.4\pm0.3$ s. This time constant, $\tau_{on}$ serves as an estimate for the persistence time of a cluster. However, another important metric is the time constant associated with $T_{off}$. The probability of $T_{off}$ split into equidistant bins is shown in the grey scatter plot. An exponential fit (grey, broken line) gives the time constant $\tau_{off} = 1.6\pm0.3$ s. Thus, clustering can be thought to occur at a high frequency, and this should be reflected in the fraction of time for which a cluster persists.
(I) The fraction of time for which a cluster persists is given by $f_{on} = \frac{T_{on}}{T_{on}+T_{off}}$. This quantity is shown as a box plot here. Boxes extend from the 25\textsuperscript{th} to the 75\textsuperscript{th} percentile, while the horizontal divider marks the median. The whiskers denote the 5\textsuperscript{th} and the 95\textsuperscript{th} percentile of the data. The mean $f_{on}$ was 0.58, with a total imaging time for each nucleus being $30$ s. The time limit of $30$ s arises out of chromatin motion which causes DNA-bound spots to drift out of the plane of imaging after that time. To summarize, the data suggests that an average cluster forms every $\sim1.5$ seconds and lasts for $\sim2.5$ seconds, providing both the frequency and lifetimes of clusters.  This data represents nuclei located about $30\:\%$ of the length from the anterior pole of the embryo.
}
\end{figure}
\begin{figure*}[h!]
\centering
\includegraphics[width=0.7\paperwidth]{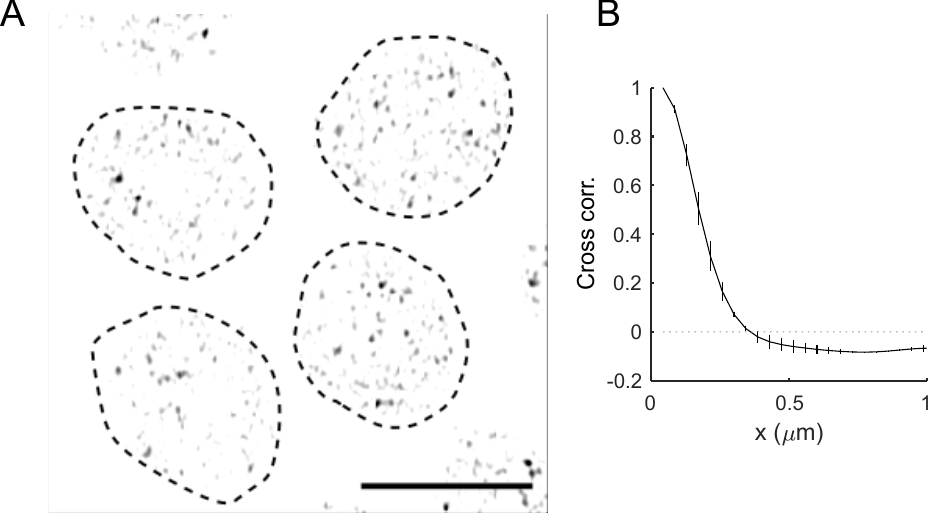}
\caption{\textbf{Bcd-GFP clusters in fixed embryos.} The goal of embryo fixation was to visualize Bcd clusters by immobilizing them in space. 
Bcd clusters form foci that are near the diffraction limit with a low signal-to-noise ratio, making it essential to avoid artifacts typically introduced by antibody staining techniques.
Therefore, we utilize the fluorescence of the monomeric eGFP tag of Bcd-GFP fusion molecules to visualize the clusters after fixation. 
This presented two challenges: 1) preserving the fluorescence of GFP after fixation, and 2) preserving the clusters themselves.
We addressed both issues by exclusively using freshly dissolved methanol-free formaldehyde (Thermo Scientific Pierce) at a final concentration of $4\:\%$ for embryo fixation. Throughout fixation and handling, we ensured that the embryos' exposure to organic solvents such as heptane, methanol, or ethanol was minimal. With these modifications to a standard protocol \cite{10.1371/journal.pbio.1000596}, fixation and visualization of Bcd clusters in the embryos were achieved. 
(A) All images and analyses shown in this manuscript are from live imaging data. However, we ensured that the clusters were also identifiable after the fixation of the embryos. The image here shows the cross-section of nuclei expressing Bcd-GFP fixed using Formaldehyde (Materials and Methods). The dashed lines are guides to the eye for the nuclear boundaries. 
The scale bar is $5\,\mu$m.
(B) Figure shows pixel correlation function (mean$\pm$s.d., n = 15 nuclei) on the nuclear pixels of a fixed Bcd-GFP embryo. Exponential fit to the plot gives a correlation length of $0.24\pm0.01\:\mu$m, which is similar to the correlation length in live embryos.
}
\label{figS1E}
\end{figure*}

\begin{figure*}[h!]
\centering
\includegraphics[width=0.8\paperwidth]{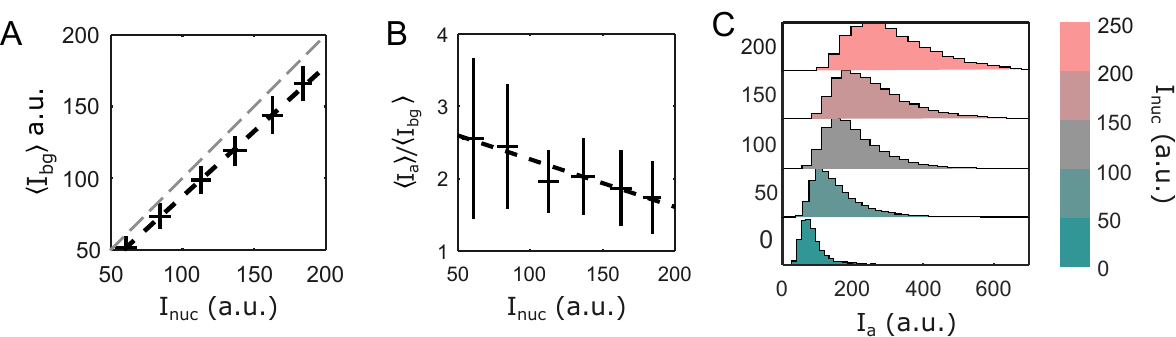}
\caption{\textbf{Molecular enrichment within clusters.} 
(A) Cluster background intensity, $I_{bg}$, as a function of average nuclear Bcd-GFP intensity ($I_{nuc}$). $I_{bg}$ is obtained from cluster fitting (Materials and Methods) and represents the intensity of Bcd-GFP at the edge of the detected clusters. Data shows $mean \pm s.d.$, fitted with a linear function (black dashed line $R^2=0.95$). The values of $I_{bg}$ are consistently lower than $I_{nuc}$ (grey dashed line representing $y=x$). This is expected since $I_{nuc}$ encompasses both $I_{bg}$ and the cluster amplitude $I_{a}$. 
(B) The nuclear average of the ratio of the cluster amplitude and the respective cluster background intensity ($I_a/I_{bg}$), plotted as a function of the average nuclear intensity $I_{nuc}$. This ratio measures the relative enrichment of Bcd molecules in clusters. It is relatively higher for lower $I_{nuc}$.
(C) Histograms showing the distribution of cluster amplitudes, $I_a$ for different average nuclear intensity bins, $I_{nuc}$. The color bar represents ranges of $I_{nuc}$.
}
\label{figS2A}
\end{figure*}

\begin{figure*}[h!]
\centering
\includegraphics[width=0.7\paperwidth]{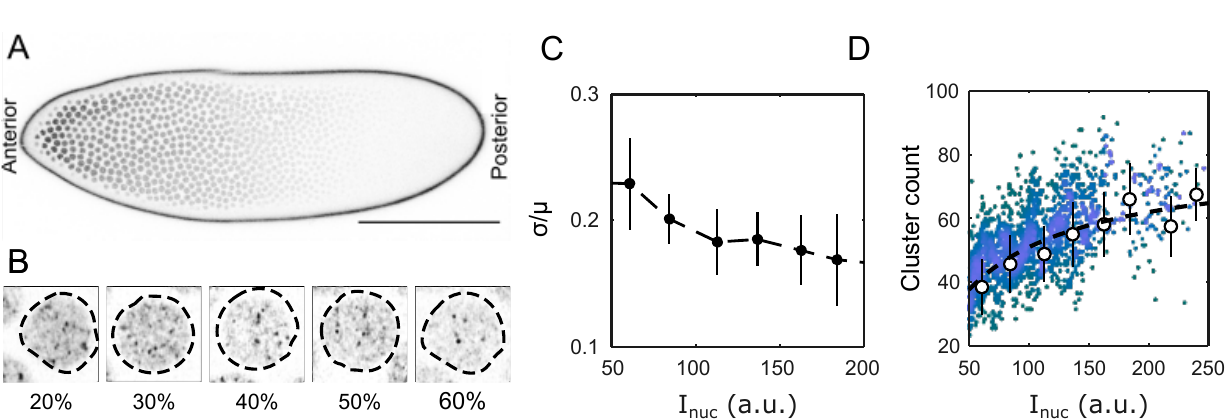}
\caption{\textbf{Cluster count.} 
(A) Cross-section of a \textit{Drosophila} embryo in nuclear cycle 14, expressing Bcd-GFP. The focal plane was set to roughly halfway between the embryo bottom and the midsaggital plane. This image, acquired using a traditional confocal setting, has a resolution much lower than the \textit{Airyscan} images. The anterior and posterior ends of the embryo are marked. The higher GFP intensities in the anterior nuclei (darker nuclei) indicate a higher concentration of Bcd.  The scale bar is $100\,\mu$m.
(B) Representative images showing cross-sections of individual nuclei expressing Bcd-GFP, imaged using the \textit{Airyscan} mode as a function of nuclear position (anterior to posterior from left to right in percentage of embryo length). The imaging settings are the same as those used for detecting and analyzing 3D clusters. Images are $8 \times 8\,\mu$m in size.
(C) Coefficient of variation (CV, $\sigma/\mu$) of the cluster count per nucleus as a function of nuclear Bcd-GFP intensity ($I_{nuc}$). Pooled data from 14 embryos, 2027 nuclei were discretized into $10$ $I_{nuc}$ bins. Mean and standard deviation were calculated for each bin using bootstrap sampling. With an average CV less than $20\:\%$, the cluster count displays remarkable reproducibility across embryos, comparable to that of the nuclear Bcd-GFP intensity ($I_{nuc}$) \cite{10.1016/j.cell.2007.05.025, 10.1073/pnas.1220912110}.
(D) Cluster count per nucleus as a function of nuclear Bcd-GFP intensity ($I_{nuc}$). Despite the high reproducibility in C, the cluster count correlates poorly with $I_{nuc}$, as also shown in Fig.~\ref{fig2}. Here we employ a simulation to understand the dependence of cluster counts on $I_{nuc}$. The scatter plot shows the number of clusters detected per nucleus. In contrast, the error bars represent the same data discretized into $10$ bins and the mean and the standard deviation of the data within these bins were calculated by bootstrap sampling. The dashed black line illustrates the simulated dependence of the cluster counts on $I_{nuc}$.
The simulation uses an empirical relation between the cluster count and $I_{nuc}$, with a model that hinges on the probability that a cluster is bound to its seeding site, denoted as  $p_{on}=t_{on}/(t_{on}+t_{off})$. 
While $t_{on}$ represents the time for which the cluster is ``bound", and hence detectable, $t_{off}$ is the time for which the cluster is ``unbound" and thus not detectable. 
To calculate these, three simple assumptions were made: 1. 
The total number of seeding sites in a nucleus is constant, denoted by $N$. 
2. The ``bound" time, $t_{on}$ for a cluster $i$ is determined solely by the properties of its seeding site, $n(i)$.
3. The ``unbound" time, $t_{off}$ depends on diffusion parameters and is inversely proportional to the nuclear concentration $c$, which correlates with $I_{nuc}$.
Therefore, the sum $\sum_{i=1}^N p_{on}(i,c)$ provides the total number of clusters detected per nucleus. We used $N=80$, which represents the maximum number of seeding sites within a nucleus in an embryo, with $n(i)$ being randomly generated. 
}
\label{figS2B}
\end{figure*}

\begin{figure*}[h!]
\centering
\includegraphics[width=0.7\paperwidth]{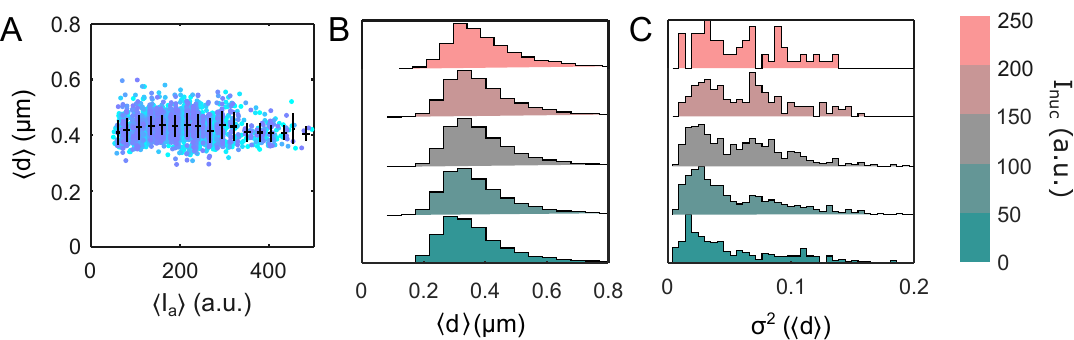}
\caption{\textbf{Cluster size.} 
(A) In Fig.~\ref{fig2}F, we showed that the cluster diameter $d$ is uncorrelated with the nuclear concentration of Bcd-GFP, $I_{nuc}$. Panel A  shows that $d$ is also uncorrelated with the cluster's Bcd-GFP concentration, $I_a$. 
(B) Cluster diameter distributions for different $I_{nuc}$ bins. The range of $I_{nuc}$ per histogram is indicated by the color bar on the right. Histograms are invariant of $I_{nuc}$ range and thus $d$ is independent of the nuclear concentration, $I_{nuc}$.
(C) Histograms of cluster size variances $\sigma ^2(d)$, binned and sorted according to the same $I_{nuc}$ ranges as in B (color as in B).
}
\label{figS2C}
\end{figure*}

\begin{figure}[h!]
\centering
\includegraphics[width=0.8\linewidth]{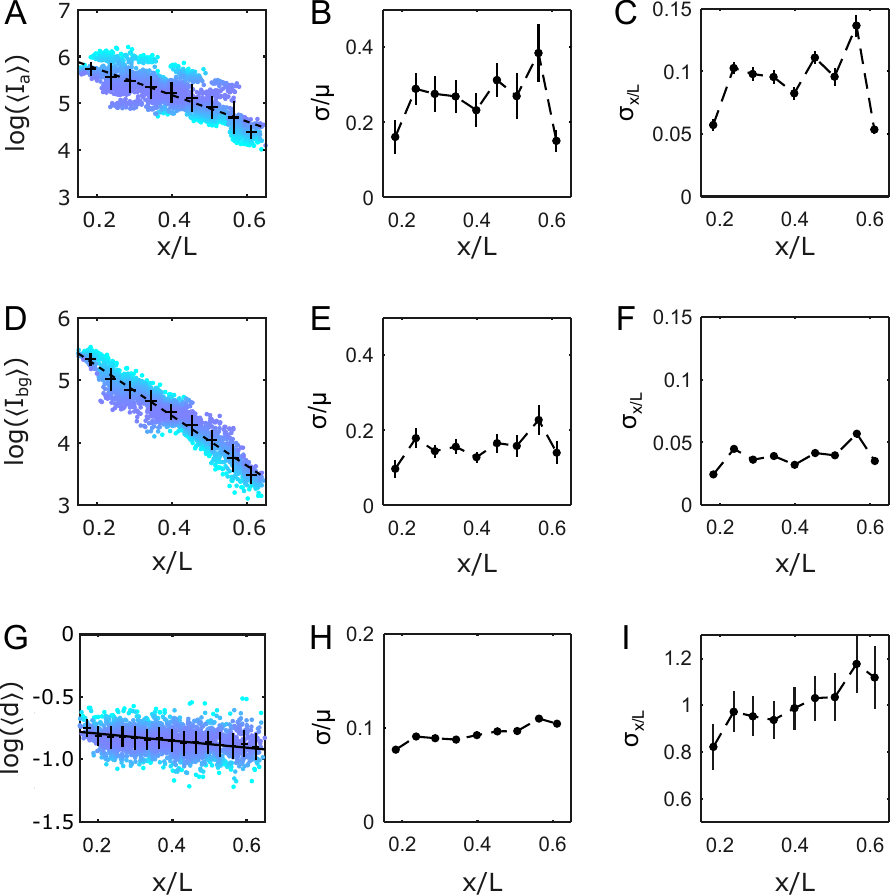}
\caption{\textbf{Precision of nuclear position determination from cluster parameters.} 
(A) Scatterplot of the nuclear average of Bcd-GFP cluster amplitudes ($\expect{I_{a}}$, log units) as a function of fractional egg length $x/L$. Data points are for individual nuclei (14 embryos, 2027 nuclei). Data is discretized into equidistant bins, and mean and s.d. (error bars) are calculated via bootstrap sampling from all nuclei within each $x/L$ bin. The exponential decay constant extracted from a linear fit (dashed line) is $\lambda=0.32\pm0.03$ EL.
(B) Coefficient of Variation (CV) of $\expect{I_{a}}$ as a function of $x/L$ in the same bins as in A. The average CV is $\sim 30\:\%$, higher than that of $I_{nuc}$ ($\sim 15\:\%$, see Fig.~\ref{fig3}B). 
(C) Figure shows the precision of nuclear position $\sigma(x)$ determined using $\expect{I_{a}}$. Utilizing $\frac{\expect{I_{a}}}{dx}$, from A and $\delta\expect{I_{a}}$ from B, we can derive $\sigma(x)$ as $\sigma(x) = \delta \expect{I_{a}} |\frac{\expect{I_{a}}}{dx}|^{-1}$. The average precision is $\sim10\:\%$, which is equivalent to $\sim4$ nuclear width. Thus $\expect{I_{a}}$ is less precise in the determination of the nuclear position than $\expect{I_c}$ ($\sim 5\:\%$, see Fig.~\ref{fig3}C).
(D) Scatterplot of the nuclear average of Bcd-GFP cluster background intensity ($\expect{I_{bg}}$, log units) as a function of $x/L$. Data is discretized into equidistant bins, and mean and s.d. (error bars) are calculated via bootstrap sampling from all nuclei within each $x/L$ bin. The exponential decay constant extracted from linear fit (dashed line) is $\lambda=0.25\pm0.02$ EL, which is similar to the gradient of $I_{nuc}$.
(E) The CV of $\expect{I_{bg}}$ expressed as a function of $x/L$. The average CV is $< 20\:\%$. Thus $\expect{I_{bg}}$ has lesser variability than $\expect{I_{a}}$ (B).
(F) The precision of nuclear position determination using the average $\expect{I_{bg}}$ is shown here. The precision is calculated as in C and is given by, $\sigma(x)\sim4\:\%$, which is equivalent to $\sim1.5$ nuclear width, similar to the precision in $I_{nuc}$. This is because $\expect{I_{bg}}$ represents the diffusing Bcd molecules which constitute $>90\:\%$ of the molecules in the nucleus. Here $\sigma(x) = \delta \expect{I_{bg}}(x) |\frac{\expect{I_{bg}}(x)}{dx}|^{-1}$, where $\frac{\expect{I_{bg}}(x)}{dx}$ is obtained from D and $\delta \expect{I_{bg}}(x)$ is obtained from E.
(G) Scatterplot of the nuclear average of Bcd-GFP cluster size, $\expect{d}$ (log units) as a function of $x/L$. Data is discretized into equidistant bins, and mean and s.d. (error bars) are calculated via bootstrap sampling from all nuclei within each $x/L$ bin. The exponential decay constant extracted from linear fit (dashed line) is $\lambda=10.7\pm0.9$ EL.
(H) The CV of $\expect{d}$ is expressed as a function of $x/L$. The average CV is $\sim10\:\%$. Thus the average size, $\expect{d}$ has the lowest variability among all cluster parameters.
(I) The precision of nuclear position determination using the average $\expect{d}$. The precision is $\sigma(x) = \sim100\:\%$, which is equivalent to $\sim1$ embryo length. This shows that $\expect{d}$ can not sense nuclear position. Even though $\expect{d}$ is a highly reproducible quantity among embryos, it fails to carry positional information due to its weak dependence on position ($x/L$). Calculation of $\sigma(x)$ follows the same logic as in C and F, giving $\sigma(x) = \delta \expect{d}(x) |\frac{\expect{d}(x)}{dx}|^{-1}$, where $\frac{\expect{d}(x)}{dx}$ is obtained from G and $\delta \expect{d}(x)$ is obtained from H.
}
\label{figS3A}
\end{figure}

\begin{figure}[b!]
\centering
\includegraphics[width=0.8\linewidth]{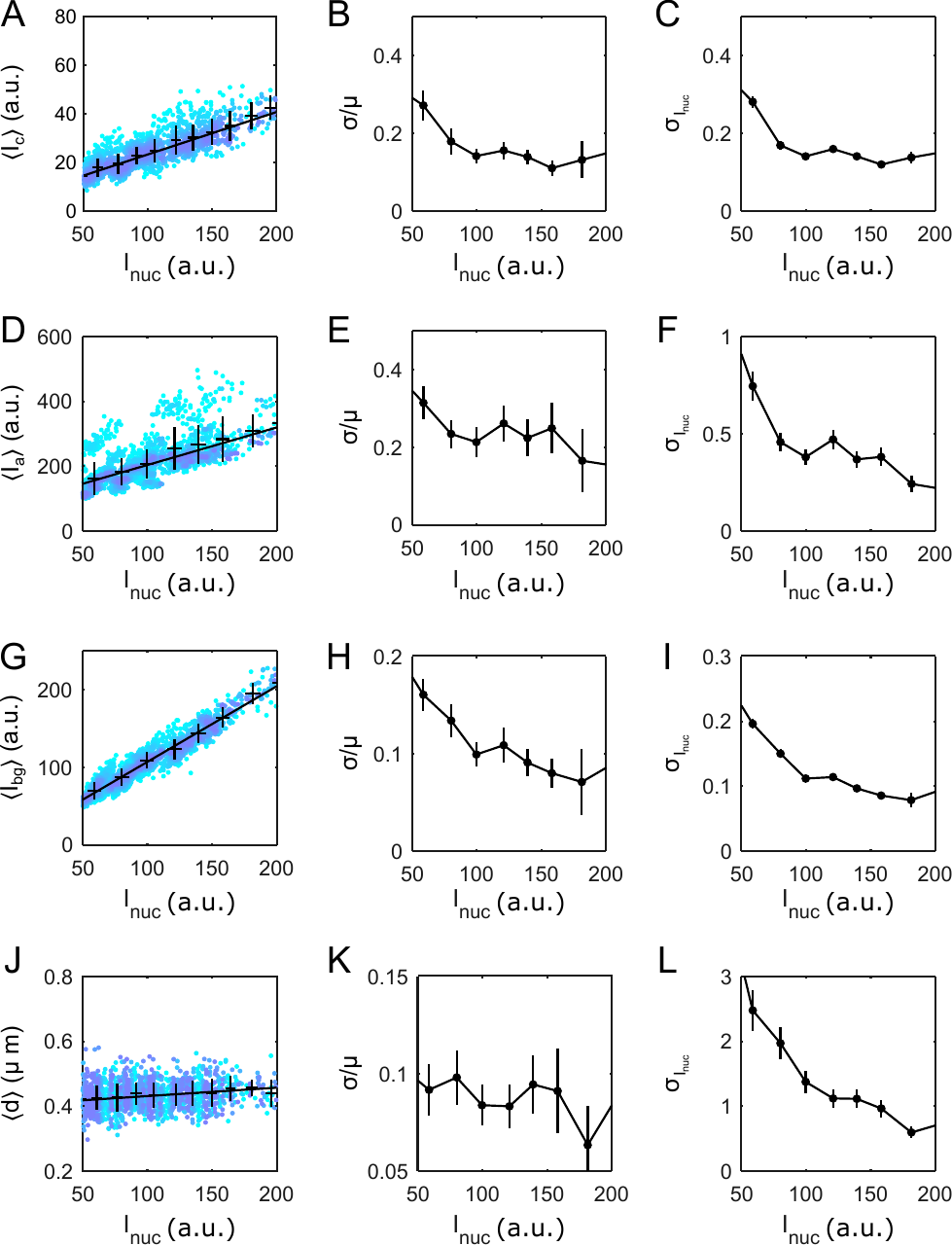}
\caption{\textbf{Precision of nuclear concentration determination from cluster parameters.} (A) Scatterplot of the nuclear average of Bcd-GFP cluster total intensity, $\expect{I_c}$ expressed as a function of nuclear Bcd-GFP intensity $I_{nuc}$. A cluster's total intensity is a measure of the total number of Bcd-GFP molecules within a cluster. Data points are for individual nuclei (14 embryos, 2027 nuclei). Data is discretized into equidistant bins, and mean and s.d. (error bars) are calculated via bootstrap sampling from all nuclei within each $x/L$ bin. The slope from the linear fit (solid line) is used to extract the precision of $I_{nuc}$ determination using $\expect{I_c}$.
(B) Coefficient of Variation (CV) of $\expect{I_c}$ expressed as a function of $I_{nuc}$. The average variability is $\sim 15\:\%$. [Caption is continued on the next page.]
}
\label{figS3B}
\end{figure}
\addtocounter{figure}{-1}
\begin{figure} [t!]
\caption{
(C) The precision of estimating $I_{nuc}$ utilizing $\expect{I_c}$ is given by $\sigma_{I_{nuc}} = \delta \expect{I_c}(c) |\frac{\expect{I_c}(c)}{dc}|^{-1}$, by combining the slope found from the linear fit in A ($\frac{\expect{I_c}}{dc}$) and the $\delta \expect{I_c}$ in B. Here, $c$ is the concentration of Bcd-GFP in a cell. The average precision of $I_{nuc}$ determination using the average $\expect{I_c}$ is $0.23\pm0.07$. Precision is relatively higher for the anterior nuclei at $\sim15\%$. Since the CV of $I_{nuc}$ is $\sim15\:\%$, $\expect{I_c}$ can be used to determine Bcd's cellular concentration $c$ with a precision of $\sim1$ cell. 
(D) Scatterplot of the nuclear average of Bcd-GFP cluster amplitude, $\expect{I_{a}}$ expressed as a function of nuclear Bcd-GFP intensity $I_{nuc}$. Data points are for individual nuclei (14 embryos, 2027 nuclei). Data is discretized into equidistant bins, and mean and s.d. (error bars) are calculated via bootstrap sampling from all nuclei within each $x/L$ bin. The slope from the linear fit (solid line) is used to extract the precision of $I_{nuc}$ determination using $\expect{I_{a}}$. The slope of the linear fit is $\frac{\expect{I_{a}}(c)}{dc}=1.19\pm0.10$ EL.
(E) The CV of $\expect{I_{a}}$ expressed as a function of $I_{nuc}$. The average CV is $\sim 25\:\%$ is higher than that of $\expect{I_c}$.
(F) The precision of estimating $I_{nuc}$ utilizing $\expect{I_{a}}$ is given by $\sigma_{I_{nuc}} = \delta\expect{I_{a}}(c) |\frac{\expect{I_{a}}(c)}{dc}|^{-1}$, by combining the slope found from the linear fit in D ($\frac{\expect{I_{a}}(c)}{dc}$) and the $\delta \expect{I_{a}}(c)$ in E. Here, $c$ is the concentration of Bcd-GFP in a cell. The average precision of $I_{nuc}$ determination using the average $\expect{I_{a}}$ is  $0.43\pm0.20$, equivalent to $\sim40\%$ error. Thus, $\expect{I_{a}}$ is less precise in the determination of the $I_{nuc}$ than $\expect{I_c}$, with a precision of $\sim3$ cells.
(G) Scatterplot of the nuclear average of Bcd-GFP cluster background intensity, $\expect{I_{bg}}$ expressed as a function of nuclear Bcd-GFP intensity $I_{nuc}$. Data points are for individual nuclei (14 embryos, 2027 nuclei). Data is discretized into equidistant bins, and mean and s.d. (error bars) are calculated via bootstrap sampling from all nuclei within each $x/L$ bin. The slope from the linear fit (solid line) is used to extract the precision of $I_{nuc}$ determination using $\expect{I_{bg}}$. The slope of the linear fit is $\frac{\expect{I_{bg}}(c)}{dc}=0.99\pm0.02$ EL.
(H) The CV of $\expect{I_{bg}}$ is expressed as a function of $I_{nuc}$. The average CV is $\sim 10\:\%$, making it a highly reproducible quantity, similar to $I_{nuc}$ (Fig.~\ref{fig3}B).
(I) The precision of estimating $I_{nuc}$ utilizing $\expect{I_{bg}}$ is given by $\sigma_{I_{nuc}} = \delta \expect{I_{bg}}(c) |\frac{\expect{I_{bg}}(c)}{dc}|^{-1}$, by combining the slope found from the linear fit in G ($\frac{\expect{I_{bg}}(c)}{dc}$) and the $\delta \expect{I_{bg}}(c)$ in H. Here, $c$ is the concentration of Bcd-GFP in a cell. The average precision of $I_{nuc}$ determination using $\expect{I_{bg}}$ is  $\sim10\:\%$ in the anterior, making the precision of concentration determination using $\expect{I_{bg}}$ as precise as a single cell. This is not surprising since $\expect{I_{bg}}$ values represent the freely diffusing Bcd molecules, which constitute $>90\:\%$ Bcd molecules in the nucleus.
(J) Scatterplot of the nuclear average of Bcd-GFP cluster size, $\expect{d}$ expressed as a function of nuclear Bcd-GFP intensity $I_{nuc}$. Data points are for individual nuclei (14 embryos, 2027 nuclei). Data is discretized into equidistant bins, and mean and s.d. (error bars) are calculated via bootstrap sampling from all nuclei within each $x/L$ bin. The slope from the linear fit (solid line) is used to extract the precision of $I_{nuc}$ determination using $\expect{d}$.
(K) The CV of $\expect{d}$ is expressed as a function of $I_{nuc}$. The average variability is $< 10\:\%$.
(L) The precision of estimating $I_{nuc}$ utilizing $\expect{d}$ is given by $\sigma_{I_{nuc}} = \delta \expect{d}(c) |\frac{\expect{d}(c)}{dc}|^{-1}$, by combining the slope found from the linear fit in J ($\frac{\expect{d}(c)}{dc}$) and the $\delta \expect{d}(c)$ in K. Here, $c$ is the concentration of Bcd-GFP in a cell.  The average precision of $I_{nuc}$ determination using the average $\expect{d}$ is $>100\%$, making cluster size an extremely imprecise metric for estimating $I_{nuc}$.
}
\label{figS3B}
\end{figure}

\begin{figure}[h!]
\centering
\includegraphics[width=0.8\linewidth]{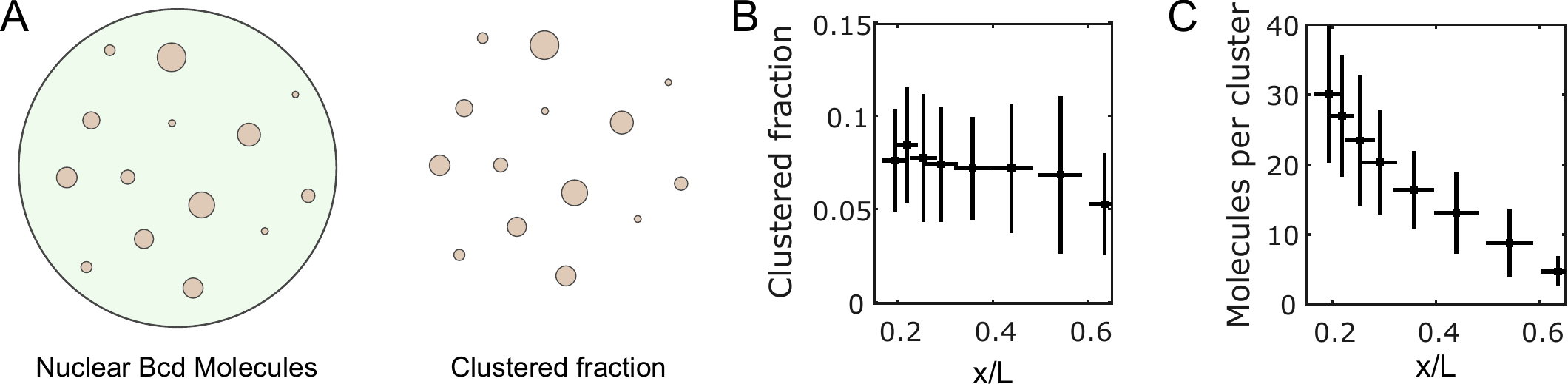}
\caption{\textbf{Estimation of the number of molecules per cluster.} 
(A) A cartoon depicting the broad distribution of Bcd molecule assemblies in the nucleus. While green depicts the freely diffusing Bcd molecules in the background, brown circles (also separately shown on the right) represent the clustered fraction. 
(B) Figure shows the fraction of Bcd molecules in the clustered fraction per nucleus as a function of the nuclear position in the embryo (error bars represent mean $\pm$ s.d. each calculated via bootstrap sampling). The cluster volume and intensity were multiplied and summed over all clusters in a given nucleus; the resulting sum was divided by the product of the nuclear intensity and the nuclear volume to obtain the fraction of Bcd molecules.
(C) The average Bcd molecule count per cluster (mean $\pm$ s.d., each calculated via bootstrap sampling)  plotted as a function of nuclear position in the embryo. The count was obtained by first calculating the absolute number of Bcd molecules in a nucleus utilizing the count obtained in \cite{10.7554/elife.28275}, and then multiplying with the fraction in B (see Materials and Methods). The average number of molecules drops from about 30 at the anterior to 5 at $60\%$ of the embryo. Note that the experimental detection limit is at $\sim5$ molecules per cluster.
}
\label{figS3C}
\end{figure}

\begin{figure}[h!]
\centering
\includegraphics[width=0.8\linewidth]{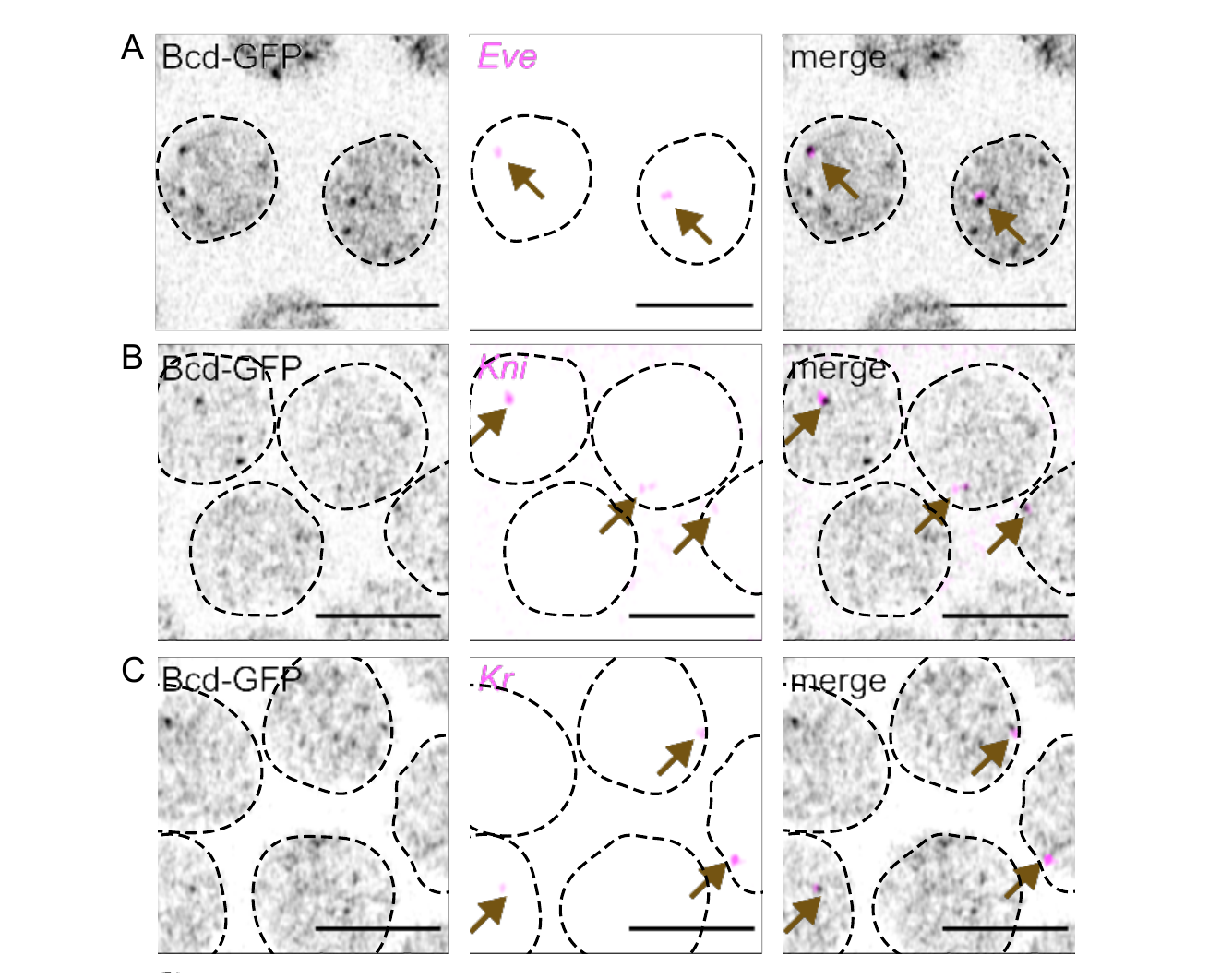}
\caption{\textbf{Colocalization of Bcd cluster with nascent transcription sites.} Representative raw images of nuclei expressing Bcd-GFP (left panels). The middle panels show the same nuclei as on the left with sites of nascent transcription of \textit{eve} (A), \textit{kni} (B), and \textit{Kr} (C). The right panels show the overlay of the left and middle images. Arrows indicate transcription sites.
}
\label{figS4A}
\end{figure}

\begin{figure}[b!]
\centering
\includegraphics[width=0.8\linewidth]{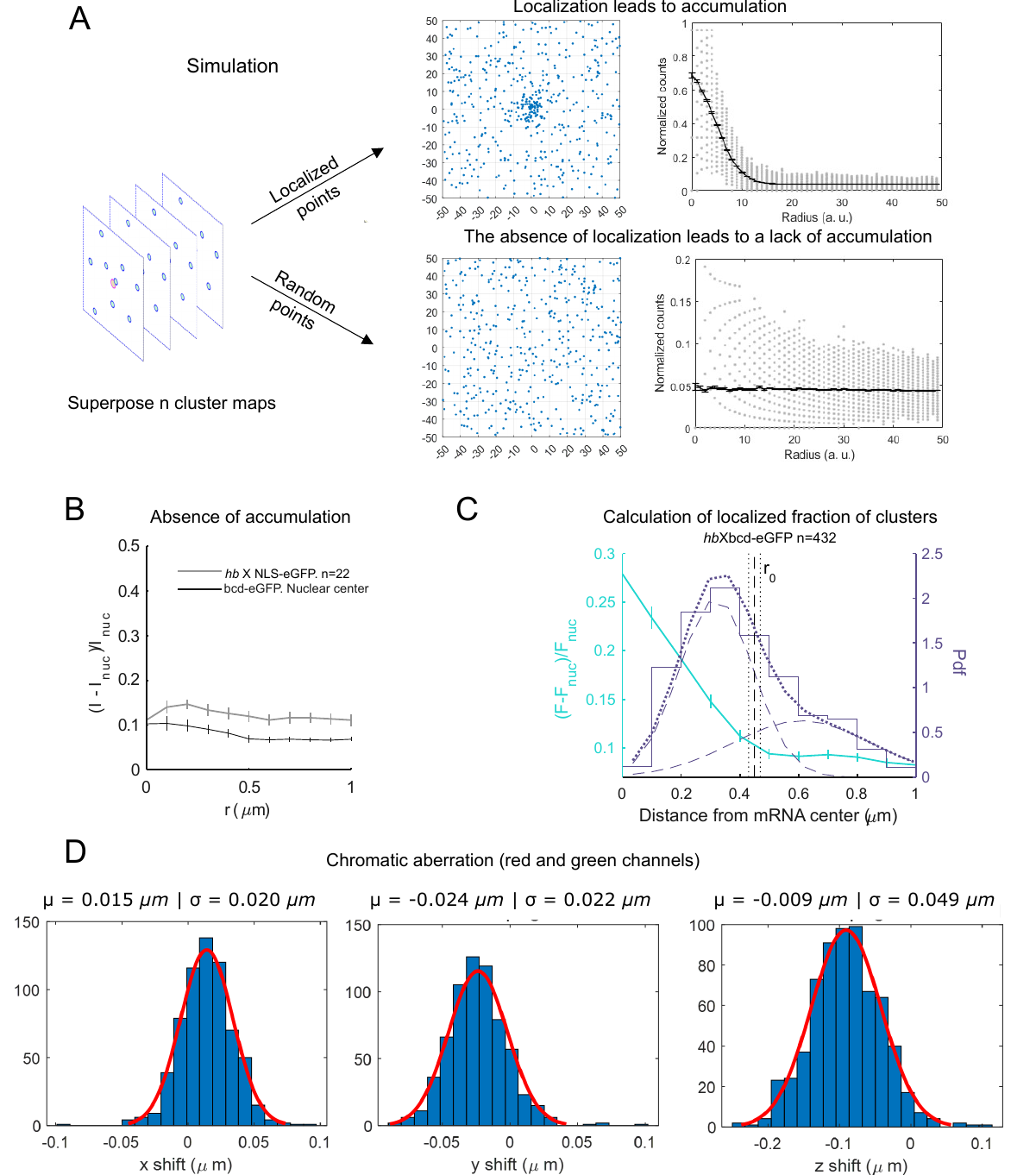}
\caption{\textbf{Bcd cluster coupling with site of nascent transcription.}
[For caption see the next page]
}
\label{figS4B}
\end{figure}

\addtocounter{figure}{-1}
\begin{figure} [t!]
\caption{\textbf{Bcd cluster coupling with site of nascent transcription.}
(A) Computer simulation of transcription factor diffusion in a space with or without a nascent transcription hotspot at its center. 
(Left) Cartoons representing a nascent transcription hotspot (pink) and TF molecules (blue). (Right, Top) Diffusing TF molecules (blue dots) at a given time are distributed as a Poisson point process across the simulation window. 
With a seeding site at the origin (emulating a nascent transcription hotspot), TF molecules preferentially accumulate there in a Gaussian-distributed fashion. This accumulation (or lack thereof) is better characterized by calculating the radial density of TF molecules, as shown on the left. 
The top plot shows the Gaussian distribution describing the accumulation of the TF molecules, the FWHM of which effectively gives the radius of the spread of the molecules around the seeding site.  Molecular clusters appearing within a radial limit given by $r_0=2\times FWHM$ are considered to be coupled to the seeding site at the origin in some capacity. 
This is elucidated in C.
(Right, Bottom) The absence of a seeding site at the origin leads to a random distribution of molecules, yielding a flat profile of the radial distribution of TF molecule density (RIGHT). Examples of this can be seen in  B.
(B) Here we show examples of the lack of accumulation of molecules, complementary to Fig.~\ref{fig4}C, showing accumulation of Bcd around target genes. For this, we imaged NLS-GFP expressing nuclei, in which a transcription site of \textit{Hb} gene was also labeled with MCD-mRuby3. 
Since \textit{Hb} is not a target gene of NLS, no accumulation was observed (gray). We could also replicate this lack of accumulation by imaging Bcd-GFP nuclei taking radial intensity profile around the nuclear center, which can be assumed to be a Bcd agnostic site (black).
(C) This plot tests the hypothesis that the molecular clusters coupled to a seeding site are preferentially found within $r_0=2\times FWHM$ of the radial accumulation of molecules around the seeding site. For this, we plot the 2D radial intensity of Bcd-GFP around a \textit{Hb} mRNA hotspot (Cyan, mean $\pm$ s.d.). 
The cyan fit gives a double Gaussian fit.  The vertical lines show mean $\pm$ standard error in the calculation of $r_0=2\times FWHM$ from the first of the two fitting Gaussians. 
In the same plot, the histogram of the distances of the clusters nearest to the \textit{Hb} mRNA hotspot is plotted. 
The nearest clusters can either be coupled (when seeded at the \textit{hb} site) or uncoupled, in which case the second-nearest cluster would be registered as the nearest one. 
To account for this, this histogram is fitted with a double Gaussian, one representing the coupled nearest cluster and the other the uncoupled nearest cluster. 
The Gaussian kernels along with the double Gaussian fit are shown in broken blue lines. The position of the intersection of the two Gaussian kernels gives the threshold distance separating the coupled from the uncoupled nearest clusters. 
This intersection is only $\sim50\:nm$ from $r_0$, obtained above, which is comparable to the dimension of a single pixel ($43$ nm). 
This warrants the use of the FWHM obtained from the radial intensity plots to find the coupling fraction of clusters in Fig.~\ref{fig4}H and Fig.~\ref{figS4C}B.
(D) Histograms of shifts in the intensity-weighted centers of polystyrene beads measured in two color channels, red and green. A Gaussian fit to each histogram gives the corresponding $\sigma$ which serves as a measure of chromatic aberration along each image axis. All three shifts are sub-pixel. 
}
\label{figS4B}
\end{figure}

\begin{figure}[h!]
\centering
\includegraphics[width=0.8\linewidth]{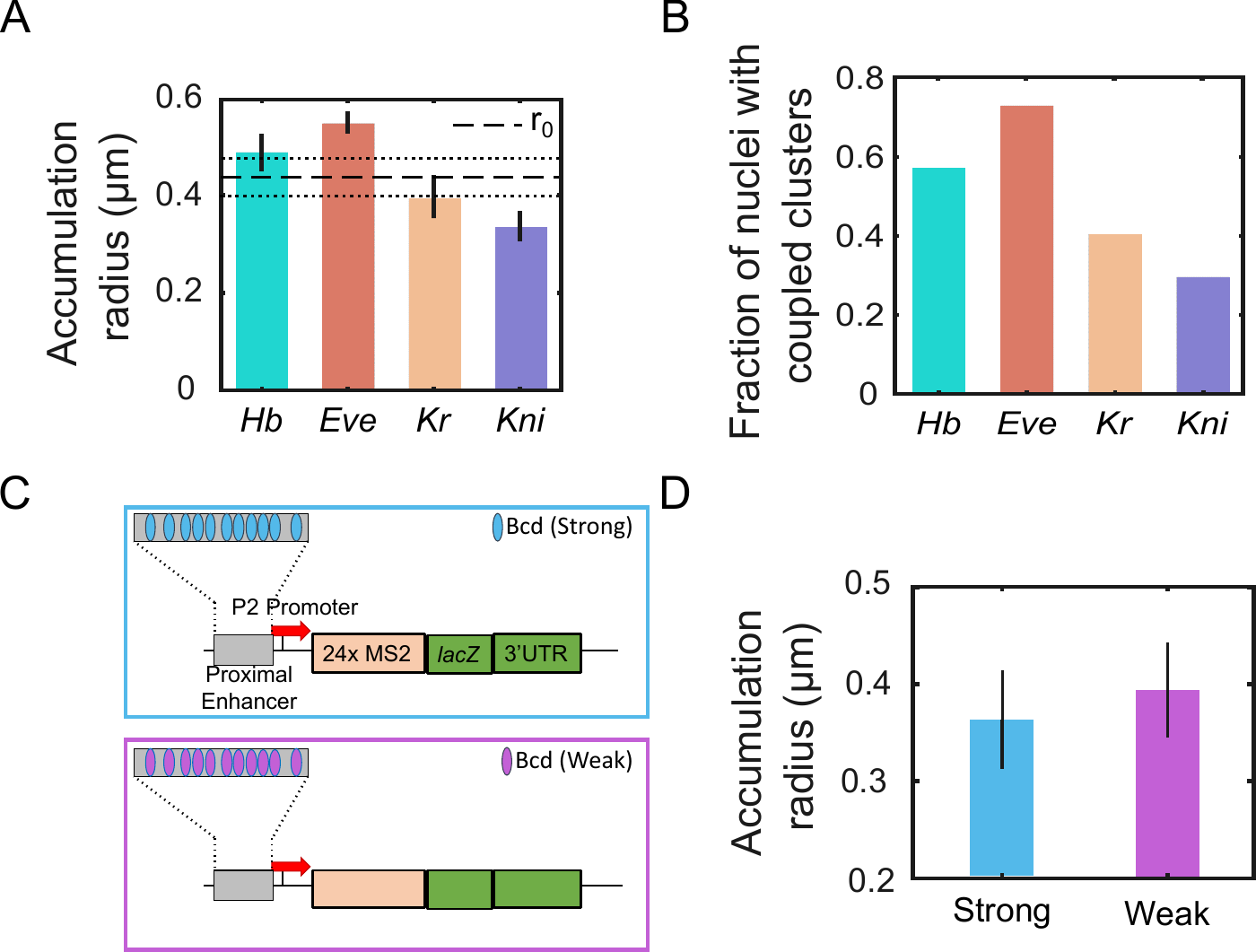}
\caption{\textbf{Fraction of transcription hotspot coupled with a Bcd cluster.} 
(A) Bars representing $2\sigma$ from the half-normal fits of Bcd-GFP accumulation (see Materials and Methods) for each target gene (indicated on the x-axis). Whiskers indicate standard errors. The horizontal line shows the average (dashed) and standard error bounds(dotted) from the entire dataset, calculated via bootstrapping. This average radius is also shown in Fig.~\ref{fig4}C by a vertical dashed line. 
(B) The fraction of nuclei in which expressing target genes are associated with a coupled cluster. The corresponding genes are indicated on the x-axis.
(C) Schematic shows the two constructs, one driven by an enhancer with strong Bcd binding sites, and the other with an enhancer composed of weak Bcd binding sites.
(D) Bars representing $2\sigma$ from the Half normal fits of Bcd-GFP accumulation for the two constructs shown in C. Following the logic in Fig.~\ref{figS4B} C, we see that the accumulation radii of both constructs are the same. This is likely because the coupled cluster location is the same for both constructs, as the enhancer is placed at the same distance from the promoter. However, in the case of the weak constructs fewer clusters are coupled Fig.~\ref{fig4}H, hence the nearest cluster distance is larger (Fig.~\ref{fig4}G) than the strong construct.
}
\label{figS4C}
\end{figure}

\begin{figure}[h!]
\centering
\includegraphics[width=0.8\linewidth]{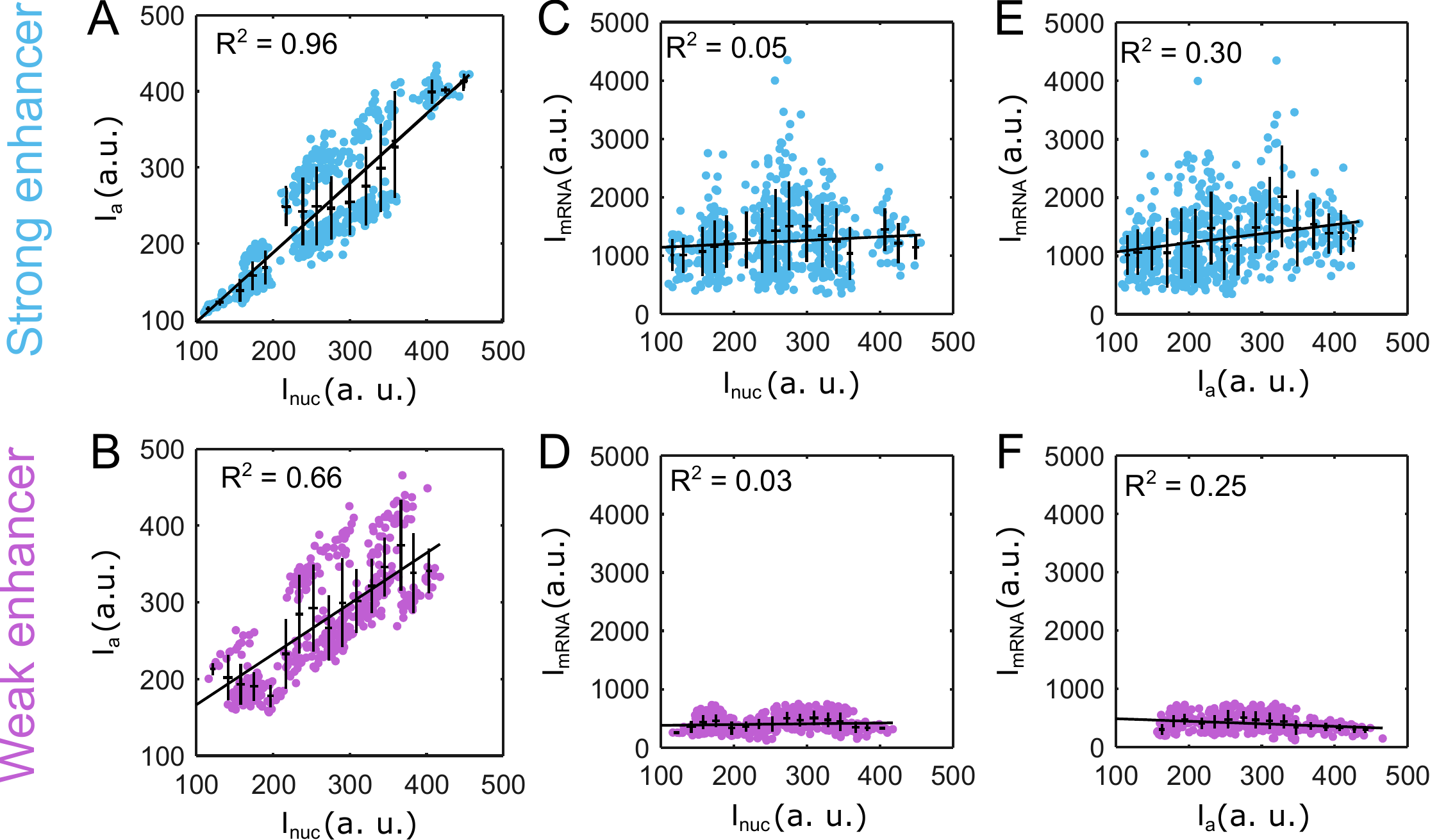}
\caption{\textbf{Dependence of cluster and transcription hotspot intensity on nuclear Bcd concentration for strong and weak synthetic enhancer constructs. } 
(A, B) Scatter plots overlaid with binned mean $\pm$ s.d. of the cluster amplitude ($I_a$) of the Bcd cluster nearest to the mRNA hotspot against the nuclear Bcd concentrations for strong (A, 541 nuclei, 17 embryos) and weak (B, 406 nuclei, 20 embryos) enhancers. The mean and standard deviations were each calculated via bootstrap sampling. Each plot is linearly fitted, and the corresponding R\textsuperscript{2} value is indicated in the figure. A very high correlation (0.85, Pearson) is observed between the nearest $I_a$ and $I_{nuc}$ for the strong enhancer (A), while a relatively weaker correlation (0.77, Pearson) is observed for the weak enhancer.
(C, D) Scatter plots overlaid with binned mean $\pm$ s.d. of the intensity of the mRNA hotspot against the nuclear Bcd concentrations for strong (C) and weak (D) enhancers. The mean and standard deviations were each calculated via bootstrap sampling. Each plot is linearly fitted, and the corresponding R\textsuperscript{2} value is indicated in the figure. The intensity of the nascent mRNA hotspot and $I_{nuc}$ are uncorrelated for both enhancer constructs.
(E, F) Scatter plots overlaid with binned means and standard deviations of the intensity of the mRNA hotspot against the intensity of the Bcd cluster nearest to the mRNA hotspot for strong (E) and weak (F) enhancers. The mean and standard deviations were each calculated via bootstrap sampling. The mean and standard deviations were each calculated via bootstrap sampling. Each plot is linearly fitted, and the corresponding R\textsuperscript{2} value is indicated in the figure. The correlations observed between the two quantities are weak for both enhancer constructs. This may be due to the significantly different characteristic persistence times of Bcd clusters and transcriptional bursts.
}
\label{figS4D}
\end{figure}

\begin{figure}[h!]
\centering
\includegraphics[width=0.5\linewidth]{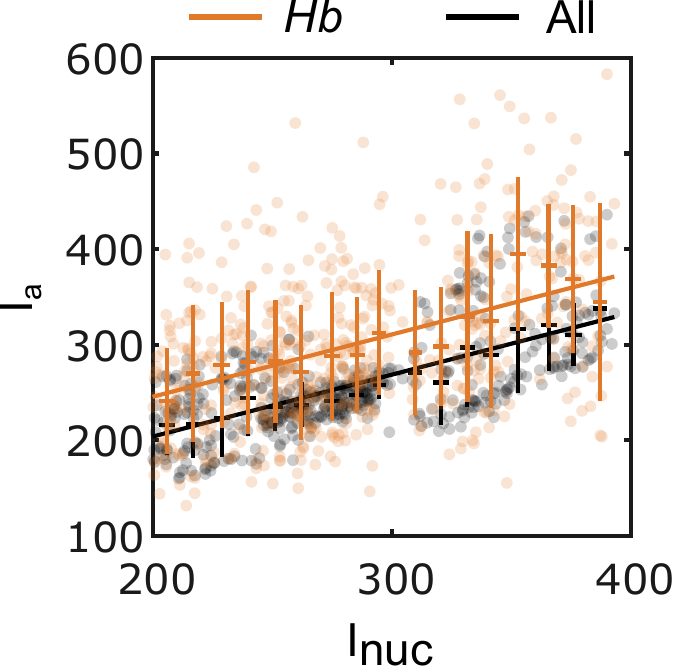}
\caption{\textbf{Cluster nearest to \textit{Hb} is brighter than the average cluster.} 
Scatter plot showing the average Bcd intensities of clusters closest to the mRNA hotspot in a \textit{Hb} reporter construct (orange); for comparison, average intensity of all clusters in a given nucleus in black (423 nuclei, 23 embryos). Overlaid are mean $\pm$ s.d. of $I_a$ calculated over equal $I_{nuc}$ bins. The linear fits, shown for data of corresponding colors (R\textsuperscript{2} = 0.93 for the nearest Bcd cluster, R\textsuperscript{2} = 0.71 for the average of all clusters) are guides to the eye. Thus, the Bcd-GFP clusters nearest to the \textit{hb} locus are brighter than an average cluster in the nucleus. 
}
\label{figS4E}
\end{figure}

\begin{figure}[h!]
\centering
\includegraphics[width=0.7\linewidth]{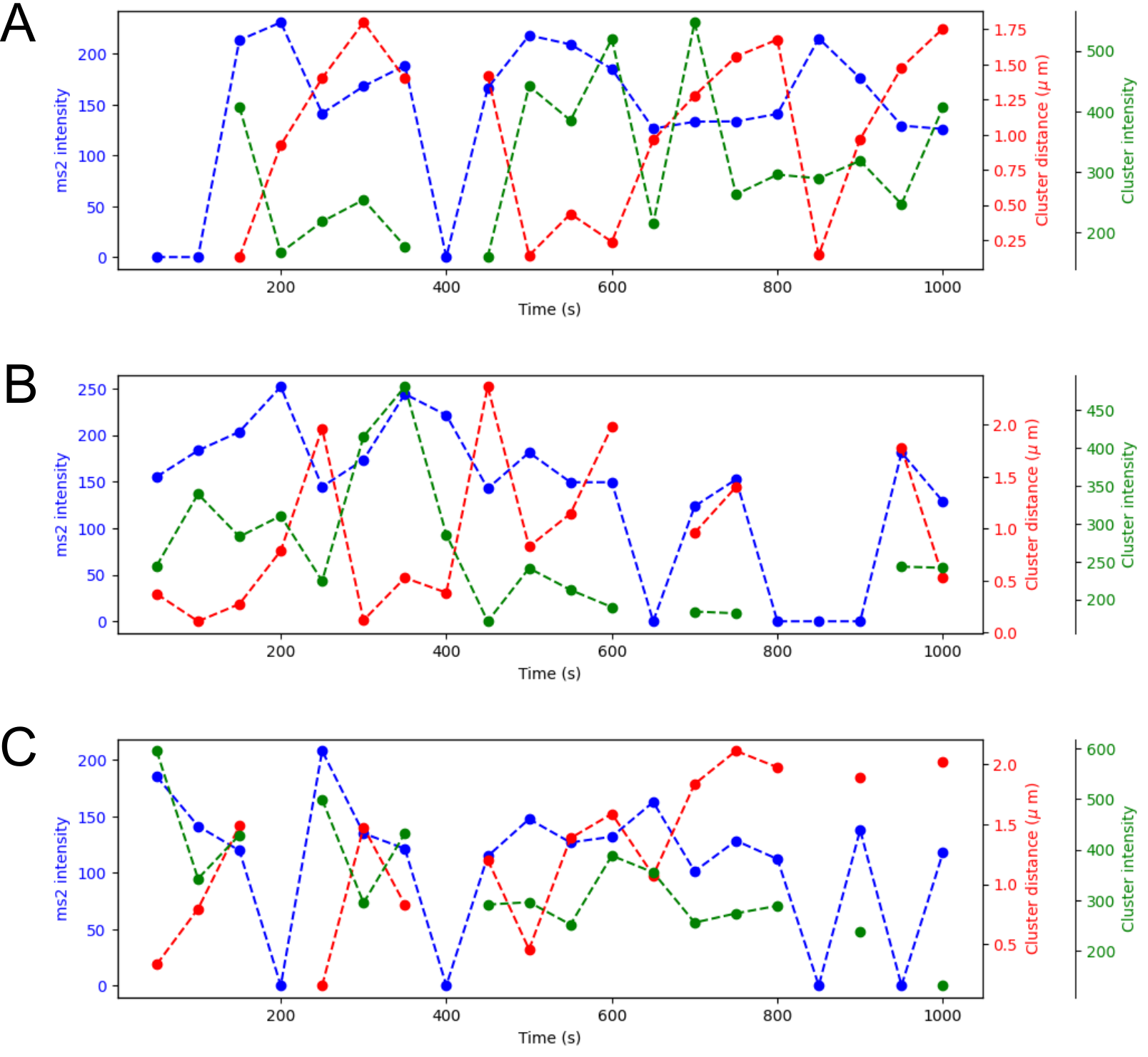}
\caption{\textbf{Nearest TF cluster distance and the mRNA hotspot intensity.} To understand how the nearest cluster affects gene transcriptional output, we imaged nuclei expressing Bcd-GFP and MCP-mRuby3 in 3D. The endogenous \textit{eve} locus was tagged with MS2 stem-loops (Materials and Methods). For each video frame, we computed the intensity of the mRNA hotspot and the relative distance and intensity of the nearest TF cluster. All imaging was done at the location of the second \textit{eve} stripe expression, during nuclear cycle 14, from 25 to 40 minutes after the end of mitosis 13.
(A, B, C) Show example traces of mRNA hotspot intensity (blue), nearest Bcd cluster distance in $\mu$m (red), and nearest Bcd cluster intensity (green) for an endogenously tagged \textit{eve} locus as a function of time. An increase in the nearest cluster distance is followed by a decrease in mRNA hotspot intensity, while no pattern is observed between the nearest cluster intensity and mRNA hotspot intensity.
}
\label{figS4F}
\end{figure}


\clearpage
\section{Supplemental tables}

\begin{table*}[h!]
\centering
\begin{tabular}{ |p{4cm}|p{4cm}|  }
\hline
Fly line & Source\\
\hline
\emph{hb} BAC$<$MS2 & Bothma \emph{et al.}\cite{10.7554/elife.07956}\\
\emph{kni} BAC$<$MS2 & Bothma \emph{et al.}\cite{10.7554/elife.07956}\\
\emph{eve} MS2 & Chen \emph{et al.}\cite{10.1038/s41588-018-0175-z}\\
\emph{kr} MS2 & El-Sherif \emph{et al.}\cite{10.1016/j.cub.2016.02.054}\\
\emph{bnk} MS2 & This work\\
\emph{P2 (Strong)} MS2 & This work\\
\emph{P2 (Weak)} MS2 & This work\\
\hline
\end{tabular}
\caption{\textbf{List of MS2 stem loop fly lines.}
}
\label{table:1}

\end{table*}

\end{document}